\documentclass[preprint,11pt, authoryear]{elsarticle}
\usepackage{amsmath}
\usepackage{amssymb}
\usepackage[a4paper, total={6in, 9in}]{geometry}
\usepackage{graphicx} 
\usepackage{subcaption} 
\usepackage{moreverb} 
\usepackage{array}
\usepackage[pdftex,colorlinks=true,urlcolor=blue,citecolor=black,anchorcolor=black,linkcolor=black]{hyperref}
\usepackage{bigstrut}
\usepackage{enumitem}
\usepackage[normalem]{ulem}
\usepackage{rotating}

\usepackage{tablefootnote}
\usepackage{array}
\usepackage{multirow}
\usepackage{booktabs}
\usepackage{tabularray}
\usepackage{lscape}
\newcolumntype{C}[1]{>{\centering\let\newline\\\arraybackslash\hspace{0pt}}m{#1}}
\newcolumntype{L}[1]{>{\raggedleft\let\newline\\\arraybackslash\hspace{0pt}}m{#1}}
\newcolumntype{R}[1]{>{\raggedright\let\newline\\\arraybackslash\hspace{0pt}}m{#1}}
\usepackage{algorithm}
\usepackage{algorithmicx}
\usepackage{algpseudocode}
\algdef{SE}[SUBALG]{Indent}{EndIndent}{}{\algorithmicend\ }%
\algtext*{Indent}
\algtext*{EndIndent}

\journal{}

\begin{document}

\begin{frontmatter}

\title{A FRAMEWORK FOR PREDICTING RUNTIME SAVINGS FROM DISCRETE-EVENT SIMULATION MODEL SIMPLIFICATION OPERATIONS}

\author[inst1]{Mohd Shoaib}

\affiliation[inst1]{organization={Department of Mechanical Engineering, IIT Delhi},
            addressline={Hauz Khas}, 
            city={New Delhi},
            postcode={110016}, 
            country={India}}

\author[inst2]{Navonil Mustafee}
\author[inst1]{Varun Ramamohan}

\affiliation[inst2]{organization={Centre for Simulation, Analytics and Modelling},
            addressline={University of Exeter}, 
            city={Rennes Drive, Exeter},
            postcode={EX4 4ST}, 
            country={United Kingdom}}

\begin{abstract}
Abstraction or substitution and aggregation are the most widely used simulation model simplification operations. Abstraction involves replacing subsystems within a discrete-event simulation (DES) with one or more quantities - typically random variables - representing the lengths of stay in the subsystems(s) in question to create a `simplified' system comprising only of subsystems of interest to the analysis at hand. Aggregation involves replacing more than one subsystem of the original `parent' simulation with a single subsystem. However, the model simplification process itself can be expensive, in terms of the computational runtime and effort required to collect the data required to estimate the distributions of the length of stay variables, the distribution-fitting process, and testing and validation of the simplified model. Moreover, the savings in simulation runtime that the simplification process yields is \textit{a priori} unknown to the modeller. In this context, a method that predicts the runtime savings (RS) from DES model simplification operations before their execution - at the conceptualisation stage of the simplified model development process - may help judge whether its development is indeed worth undertaking. In this paper, we present a queueing-theoretic framework for the prediction of RS from model simplification operations. Our framework is applicable for DES models comprising $M/M/, M/G/ \text{ and } G/G/$ subsystems. The performance of the RS prediction framework is demonstrated using multiple computational experiments. Our proposed framework contributes to the literature around DES model complexity and more broadly to DES runtime prediction.


\end{abstract}

\end{frontmatter}


\section{Introduction}
\label{sec:intro}

A discrete event simulation (DES) model designed for large and complex systems may be \textit{simplified} for a more focused performance analysis involving a subset of the system. For example, a detailed hospital simulation model, that covers all its operations, is not needed to analyse the performance of a specific department within the hospital. In such cases, a simplified version of the simulation model can be developed. This approach can significantly reduce the computational effort needed for the analysis, making it both more efficient and focused.

The practice of developing simplified simulation models is termed model simplification, and it is used as a strategy to reduce model complexity and improve computational performance \citep{rank2016correct, vanderZeeWSC}. It presents several notable advantages, including a rapid model development cycle, ease of understanding and validation, as well as reduced data requirements, among other benefits \citep{robinson2023exploring,Lidberg2021}. Multiple studies in the literature have reported significant improvements in model computational performance after simplification. For instance, \cite{Fatma2020} achieved approximately $80$\% reduction in computational runtime with simplified models. 

Typically, aggregation, deletion and abstraction or substitution approaches are employed to simplify simulation models \citep{vanderZeeWSC}. These strategies aim at reducing model elements, components, or subsystems in different ways, without altering the overall behaviour of the model. The aggregation approach involves combining multiple processes, pathways, or components together into a single and unified corresponding entity~\citep{Lidberg2021}. The deletion technique, also termed omission, is concerned with the removal of non-relevant components from the detailed model. 
The abstraction model simplification approach is based on the strategy of replacing specific subsystems within the full-featured `parent' model with simpler or less complex subsystem models. Subsystem replacement can be done using constant-valued delays or random variables in place of the subsystem being abstracted out. These delays could represent cycle times of different machines \citep{Brooks2000} or lengths of stay in the subsystem \citep{Fatma2020}. In this paper, we are interested in the broadly applicable model simplification operations abstraction (substitution) and aggregation. More specifically, we are interested in the application of these model simplification operations to DESs of complex queueing systems which comprise $M/M/n, M/G/n, \text{ or } G/G/n$ subsystems or workstations.

\begin{figure}[htbp]
    \centering
    \includegraphics[width=0.70\textwidth]{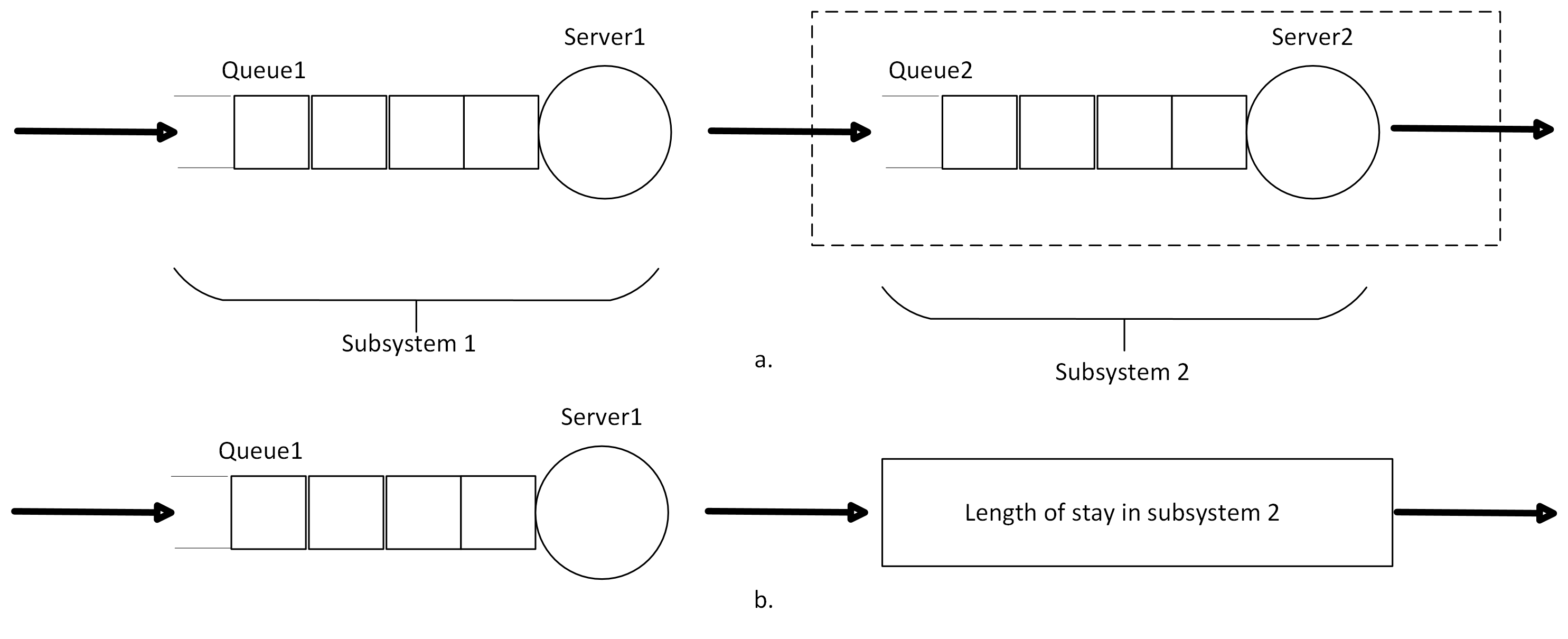}
    \caption{Top: a two-stage tandem queuing system (parent model). Bottom: A simplified version model of the parent model developed by substituting the second subsystem with a random variable representing the length of stay in the second subsystem.}
    \label{figure:abstraction}
\end{figure}

The abstraction operation can be illustrated using a simple 2-stage tandem queuing system depicted in Figure \ref{figure:abstraction}. The top half of Figure \ref{figure:abstraction} shows the parent model composed of two subsystems, subsystem-1 and subsystem-2 in series, whereas the bottom half of the figure depicts the simplified version. Considering that we are interested in the following: a) the performance analysis of subsystem-1; b) the overall length of stay (LOS) in the system, and c) reducing the computational runtime of the model, without changing the overall stochastic behaviour of the model. This could be accomplished by constructing a simplified model, shown in the bottom half of \ref{figure:abstraction}, wherein subsystem-2 is abstracted out using a random variable representing the LOS in the subsystem. Such a simplified model retains the stochastic character of the system and improves the computational costs associated with model execution.
Moreover, it is worth mentioning that the terms `abstraction' and `substitution' are often used synonymously in the literature, but in this paper we will use the term `abstraction'. 

\begin{figure}[htbp]
    \centering
    \includegraphics[width=0.60\textwidth]{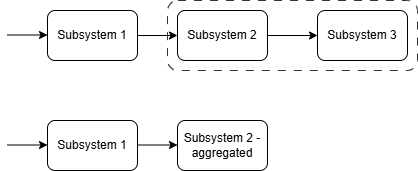}
    \caption{Top: a three-stage tandem queuing system (parent model). Bottom: a simplified version of the parent model developed by aggregating the second and third subsystems into a single subsystem.}
    \label{fig:aggregation}
\end{figure}

Similarly, the aggregation operation is depicted in Figure~\ref{fig:aggregation}, wherein the queueing system in the top half is a tandem queueing system with three subsystems. The dashed box around the second and third subsystems indicates that these subsystems will be subject to an aggregation operation. The queueing system in the bottom half is the simplified model, wherein the second and third subsystems in the parent model are replaced by a single subsystem with characteristics pertaining to the systems subject to the simplification operation - for example, the LOS of the second subsystem in the simplified model will be the same as that of an entity in the aggregated subsystems in the parent model.

The question of whether computational runtime savings (RS) can be predicted during the model conceptualisation phase has not, in our knowledge, been  addressed in the literature. Additionally, the existing literature does not in our knowledge address the mechanisms and determinants of RS gained from DES model simplification operations. 
In this work, we develop RS prediction models for DESs composed of $M/M/$, $M/G/$ and $G/G/$ systems, implying that our proposed methodology can be applied to predict RS at the conceptualisation stage for a large subset of commonly developed DES models. Our predictive models can be used to estimate the expected RS before an abstraction or aggregation operation is performed. This allows for informed decision-making regarding whether to proceed with the simplification process at all. Further, we have tested the performance of the RS prediction framework through a set of computational experiments across systems with varying complexities, including a real-world case study. We provide a more detailed description of the research contributions of this study in the following section on related work.





    
The remainder of the paper is organised as follows. Section~\ref{sec:litrev} provides an overview of the relevant literature and the research contributions of this study. In Section~\ref{sec:method}, we present the determinants of and mechanism underlying the RS achieved with model simplification, which is followed by a description of the proposed RS prediction framework. The numerical experiments conducted to develop the models within the framework and the deployment of the framework are described in Section~\ref{sec:implem}. Section~\ref{sec:valid} presents the simulation experiments conducted for assessing the performance of the RS prediction framework. We conclude in Section \ref{sec:conc} with a summary of the study and its potential impact, a discussion of its limitations and avenues for future research.

\section{Relevant Literature}
\label{sec:litrev}

Literature analysis highlights that model simplification in simulation modelling has been in use for about $30$ years. The importance of creating simple models was emphasised in the work of \cite{Brooks2000} where authors argued that complex models including too many details often lead to project failure.

\cite{huber2009controlled} define model simplification as the process of reducing complexity to decrease the runtime and computing resources needed to execute a simulation model. 
It must be noted that there is a wide variation in the understanding of model complexity. Some authors link complexity to the computing power and execution time required to run the model \citep{huber2009controlled, schruben1993complexity}. In contrast, others propose that it is the number of components, connections, and computations in a model measure of its complexity \citep{Brooks2000}. Despite the divergence, a general consensus exists that model simplification reduces complexity and leads to more computationally efficient models.  
Thus, it could be argued that model simplification is undertaken to reduce a simulation model's complexity and computational costs associated with model execution. For example, \cite{Fatma2020} created a simplified healthcare facility network model to reduce the computational time needed to execute the detailed model.  Similarly, a DES model of a wafer fabrication facility was simplified by ~\cite{Hung1999} to improve the model runtime performance. The simplified model achieved approximately 20\% reduction in execution time in comparison to the detailed model.      

Large-scale and complex simulation models, as encountered in the fields of semiconductor manufacturing and healthcare among others, are computationally expensive to run \citep{adan2014aggregate, Fatma2020}. Moreover, as the number of replications required to reduce the output variance and improve overall output reliability increases, the computational overheads increase considerably. In such cases scenario analysis and optimisation in particularly become challenging to execute. To overcome the computational requirements, modellers may resort to model simplification via different strategies.
Among these strategies abstraction and aggregation have become the most adopted approaches \citep{rank2016correct, pegden1995introduction, Lidberg2021}.

Abstraction, substitution, or model condensation \cite{Piplani2004}, entails substituting non-essential elements from a comprehensive or detailed model. A prevalent convention in the literature is substituting particular components or elements with constant delays \citep{Hung1999, johnson} or with random variables representing delays in the substituted components \citep{Etman2011, Fatma2020}. 
Aggregation for model simplification is a technique where multiple process flows, components, elements, or machines are combined to form a single unified compound process or machine \citep{Lidberg2021}. The primary objective of aggregation is to simplify models by combining flows and machines to facilitate faster model execution and experimentation. Several simulation elements, for instance, can be represented by a single unified element, and a single random variable can represent processing time. Deletion refers to reducing the model by deleting/removing the non-essential components from a model.  


\underline{\textit{Studies Reporting Model Simplification Endeavours.}}
Our literature review revealed several examples of simulation model simplification using abstraction approach. These include replacement of non-bottleneck workstations \citep{Rose1998,Rose1999,Rose2007, Hung1999}, tool-sets \citep{Stogniy2019,Stogniy2020}, and subsystems not central to the analysis at hand \citep{Fatma2020,pehrsson2015aggregated,volker}. In terms of application areas, it can be observed that the literature is replete with instances from manufacturing sector, and specifically semiconductor fabrication \citep{Rose1998, Rose1999, roy2020application, Stogniy2019, Stogniy2020}. A probable explanation for this could be found in the fact that these systems are inherently complex and often require rapid decision-making necessitating what-if scenario analysis. Given the complexity of the system and time scales in decision-making, experimentation with a detailed model becomes challenging due to higher associated computational costs; hence, model simplification is performed. 
Model simplification in semiconductor manufacturing system analysis could be found in the work of \cite{Rose1998, Rose1999, Rose2007}. In \cite{Rose1998}, the authors developed a simplified simulation model to analyse the behaviour of a wafer production plant. In their approach, the author included only the bottleneck machine in its full-featured format while replacing the other machines with a combined delay (random variable) corresponding to all the abstracted machines. The author utilised the simplified model to examine work-in-process levels following a catastrophic bottleneck failure. 
In a follow-up study \citep{Rose1999}, the authors performed a statistical analysis of the lot interarrival time for the bottleneck workstation using the simplified model. To accomplish the task they determined the most suitable probability distribution to accurately represent the interarrival time data, and then used it to model the arrivals at the bottleneck machine. The use of the simplified model was further extended to a subsequent study \citep{Rose2007}. For this, the authors introduced inventory-dependent distributions for sampling the delays as opposed to the fixed distributions employed in the previous research.  
\cite{Hung1999} demonstrated that simplified models maintain the accuracy of the detailed models while significantly reducing the computer execution time. In their experiment, the authors substituted underused or low-utilisation workstations with fixed and random delays, representing the durations of lot wait and processing times. The authors reported the behaviour of mean error under different scenarios and gained a reduction of up to 90\% in the model execution time.
In healthcare settings, \cite{Fatma2020} developed simplified simulation models of a network of primary care institutions. The authors focused solely on the components pertinent to their research and substituted the non-essential components/subsystems with random variables representing the duration of patient stays (which includes both wait time and service time) at each subsystem. The authors also performed aggregation of servers at a workstation by combining multiple servers into one while maintaining overall workstation occupancy. The authors reported a maximum reduction of 80\% in model runtime.

It is evident from the above discussion that reduction in computational runtime is an important reason for simplifying DES models. However, the extent of RS achieved remains unavailable to modellers until a model has been simplified and then executed. Without prior knowledge of the expected RS, it becomes difficult to make informed decisions regarding the extent and necessity of model simplification.
Notably, our review did not identify any studies that attempted to \textit{a priori} predict the RS that may be achieved via abstraction or aggregation - that is, at the conceptualisation stage of developing the simplified model. Therefore, our primary research contribution involves addressing this gap, building upon preliminary research on the development and application of an RS prediction appproach \citep{shoaibWSC2023}. \cite{shoaibWSC2023} develop an approach for predicting RS from abstraction simplification operations conducted for DESs comprised of $M/M/n$ subsystems and provide a brief demonstration of its application for DESs of simple queueing systems. This study extends the work in \cite{shoaibWSC2023} in the following ways: (a) first, we consider DESs comprising $M/G/n$ and $G/G/n$ subsystems in addition to $M/M/n$ subsystems, which are much more widely encountered; (b) we also show how our proposed RS prediction framework can be applied for aggregation (in addition to abstraction) simplification operations; (c) we provide a formal description of how the RS predictive framework can be deployed for abstraction and aggregation operations; and (d) provide a much more comprehensive set of computational experiments to validate the predictive framework, including its application to a DES of a real-world system. 

\underline{\textit{DES Runtime Prediction: Brief Overview.}}
As will be described in the subsequent section, our proposed RS predictive framework involves models that first predict the number of instructions executed by the processor per arrival to the subsystem being abstracted out as a function of its occupancy, and then predicts RS as a function of the reduction in instructions achieved by the simplification operation. The number of instructions per arrival associated with a subsystem is closely linked to its simulation runtime (per arrival). Therefore, even though our study does not directly attempt to predict subsystem simulation runtime, our proposed framework contributes to the field of simulation runtime analysis and prediction.

Generating (accurate) simulation runtime predictions in a practical setting is a nontrivial task, in part because simulation runtime depends, in addition to model complexity, to a large extent on the computational resources available and the state of the computer system on which it is being executed (e.g., the number of other tasks running). In our knowledge, it appears that the majority of previous work on DES performance prediction appears to have been done for parallel and distributed asynchronous and synchronous DESs, where the goal is to predict the speed-up, or bounds on speed-up achieved through various simulation protocols \citep{cheng1990state}. The methodologies adopted include linear programming based event scheduling across multiple CPUs \citep{kunz2011predicting}, graph-theoretic critical path methods that examine event graphs \citep{lim1999performance}, and communication models that also consider computation granularity \citep{xu2004predicting}. Trace-based analyses to perform synchronisation of parallel simulations by reconstructing event dependencies have been done \citep{bagrodia2000performance}, including trace-based performance analysers that explicitly take hardware characteristics into account \citep{juhasz2003performance}. However, in our search of this area of the literature, we could not find another study that attempted to directly predict asynchronous DES runtime by via a queueing-theoretic approach such as ours that predicts the number of instructions as a function of server occupancy. This also thus forms a contribution of our study.


\section{Runtime Savings Prediction Methodology}
\label{sec:method}

We start by determining the mechanism underpinning RS in model simplification operations. Our investigation of the mechanism by which RS are generated provided the key insights that led to the development of the RS prediction methodology. We must mention here that much of our proposed methodology was derived by examining outputs from DESs developed and executed on the open-source Python library \textit{salabim} for DES \citep{van2018}. A similar exercise may need to be carried out for other platforms. The Python code used in the development and validation of the RS models is available on our \href{https://github.com/shoaibiocl/ArrivalCount-2-stage-System-}{GitHub} page. Additional details are provided in \ref{app:github}.

\subsection{Runtime Savings Mechanism Associated with Model Simplification Operations}
\label{crr_method}
We first introduce the terminology used in the development of the RS prediction methodology. 
\begin{itemize}
    \item The `event list' refers to the sequential record of all future events scheduled to be executed.
    \item The `simulation trace', on the other hand, is a log of events that have already occurred, organised chronologically. The trace is often found as one of the default metadata outputs in many simulation packages, including \textit{salabim}.
    \item The `generator' is a class that when instantiated generates simulation entities or objects (e.g., jobs, patients) that undergo some form of processing in a DES model.
    \item The term `instruction' denotes a command or message relayed by the simulation engine to execute events recorded on the event list. A specific set of instructions is associated with the execution of each event: for example, to generate a simulation entity, the instruction executed by the engine will be `\textit{create entity}'. Similarly, the instruction to enter a queue would be `\textit{enter queue}'.
    \item  For the purposes of this paper, we also define a `subsystem' as a combination of a queue and one or several servers within a DES model. This definition is based on the understanding that the majority of DES models are composed of one or many such subsystems.
\end{itemize}


We now describe the process of selection of independent variables for the RS prediction models. A DES progresses through a sequence of events that occur sequentially over time. A set of instructions is executed by the simulation engine to process these events; and as the number of events grows or the size of event list increases, the number of instructions executed by the processor also increases. 
This in turn affects the model computational runtime.
Thus, the simulation runtime may be considered as a function of the total number of instructions executed by the processor. The total number of instructions processed depends on the following three factors: 
\begin{enumerate}
    \item Number of subsystems within the model.
    \item Occupancy levels of the servers within these subsystems.
    \item Simulation run length.
\end{enumerate}

For a constant simulation run length, computational runtime depends on the number of subsystems in the model and the individual subsystem server occupancy levels.  
With regard to the number of subsystems, consider a scenario wherein a subsystem is added to an existing DES model. The simulation engine will execute additional instructions associated with the events of the newly added subsystem; consequently, a larger set of instructions will need to be executed, in turn increasing the simulation runtime. 
 
With respect to server occupancy, consider a case where a server is idle. When an entity arrives to the system it will then enter into service immediately. Associated with this event, the simulation engine executes a certain minimum number of instructions to, for example, generate the entity, deliver service and when the service is finished, leave the system. This scenario is more likely to occur when server occupancy is low. When an arriving entity finds a server busy - which is more likely at high server occupancy levels - it will have to join the queue, wait for its turn, and when the server becomes available leave the queue for service. Therefore, as server occupancy changes, the average number of instructions for each arrival varies accordingly; consequently, we selected server occupancy value as the determinant of the average number of instructions per arriving entity when the number of subsystems is fixed. 

To establish a relationship between server occupancy and the number of instructions, we constructed two simple simulation models: a) a two-stage tandem queueing system with both $M/M/1$ or $M/G/1$ or $G/G/1$ queueing systems (similar to that in Figure~\ref{fig:1}a), as applicable, and b) a simplified version of this model (Figure~\ref{fig:1}b), with the second stage abstracted out in lieu of a random variable capturing the LOS in the subsystem. Both simulation models were executed under identical conditions, and their execution traces were recorded (Table \ref{tab:trace}). These trace outputs detail the instructions executed by the simulation engine from the point of generation of a simulation entity to its departure from the system. It is evident that the number of instructions executed per arrival for the simplified model was significantly lesser than the number executed for the parent model. This reduction in the number of instructions executed for each arrival (RIEs) by the simulation engine for the simplified model presents a plausible explanation for the observed computational runtime savings, and hence, we selected RIEs for predicting the expected RS achieved by model simplification via abstraction. 

\begin{table}[h]
  \centering
  \caption{Example instructions from a two-stage $M/M/1$ tandem queueing system and corresponding simplified model with abstracted second stage. \textit{Notes.} $\sim$ indicates `sample from distribution'. $\beta$ = mean arrival rate. $\gamma, \gamma^\prime$ = mean service rates of servers 1 and 2, respectively. $s_\ast$ = length of stay in subsystem 2. $^{a}$Server busy, simulation component awaits its turn. $^{b}$Random variable representing the time spent in abstracted out subsystem 2.}  
  \footnotesize
  \renewcommand{\arraystretch}{1}
    \begin{tabular}{|p{7.1cm}|p{7.1cm}|}
    \hline
    \textbf{Parent model} & \textbf{Child model} \\
    \hline
    Component generator: & Component generator: \\
    1. Current & {1. Current} \\
    2. Pause for $\sim$ $exp\left(\frac{1}{\beta}\right) $ time & 2. Pause for $\sim$ $exp\left(\frac{1}{\beta}\right) $ time \\
     & \\
    \hline
    Component process: & Component process: \\
    1. Create & 1. Create \\
    2. Activate & 2. Activate \\
    3. Current & 3. Current \\
    4. Enter \textit{queue}1 & 4. Enter \textit{queue}1 \\
    5. Request \textit{server}1 & 5. Request \textit{server}1 \\
    6. Request for \textit{server}1 is scheduled for time $=\infty$, goes to passive state$^{a}$ & 6. Request for the \textit{server}1 is scheduled for time $ =\infty$, goes to passive state$^{a}$ \\
    7. Request honoured, scheduled for time $=s$ & 7. Request honoured, scheduled for time $=s$ \\
    8. At time $ = s$, current (from passive state) & 8. At time $=s$, current (from passive state) \\
    9. Leave \textit{queue}1 & 9. Leave \textit{queue}1 \\
    10. Claims \textit{server}1 and pauses for $\sim$ $exp\left(\frac{1}{\gamma}\right)$ & 10. Claims \textit{server}1 and pauses for $\sim$ $exp\left(\frac{1}{\gamma}\right)$ \\
    11. At time $s_{1}=s + \sim exp\left(\frac{1}{\gamma}\right)$ , current & 11. At time $s_1=s+\sim exp\left(\frac{1}{\gamma}\right)$, current \\
    12. Release \textit{server}1 & 12. Release \textit{server}1 \\
    13. Enter \textit{queue}2 & 13. Schedule for time $s= s_1 + \sim s_{\ast}$$^{b}$ \\
    14. Request \textit{server}2 & 14. At time $s= s_1 + \sim s_{\ast}$, current \\
    15. Request for the \textit{server}2 - scheduled for time$=\infty$, goes to passive state$^{a}$ & 15. End \\
    16. Request honoured, scheduled for time $s=s_2$ &  \\
    17. At time $s= s_2$, current &  \\
    18. Claims server and pauses for $\sim exp\left(\frac{1}{\gamma^{\prime}}\right)$ &  \\
    19. At time $s_{3}=s_2+\sim exp\left(\frac{1}{\gamma^{\prime}}\right)$, current &  \\
    20. Release \textit{server}2 &  \\
    21. End & \\ \hline   
  \end{tabular}%
  \label{tab:trace}%
\end{table}%
\normalsize

\begin{figure}[htbp]
    \centering
    \includegraphics[scale=0.7]{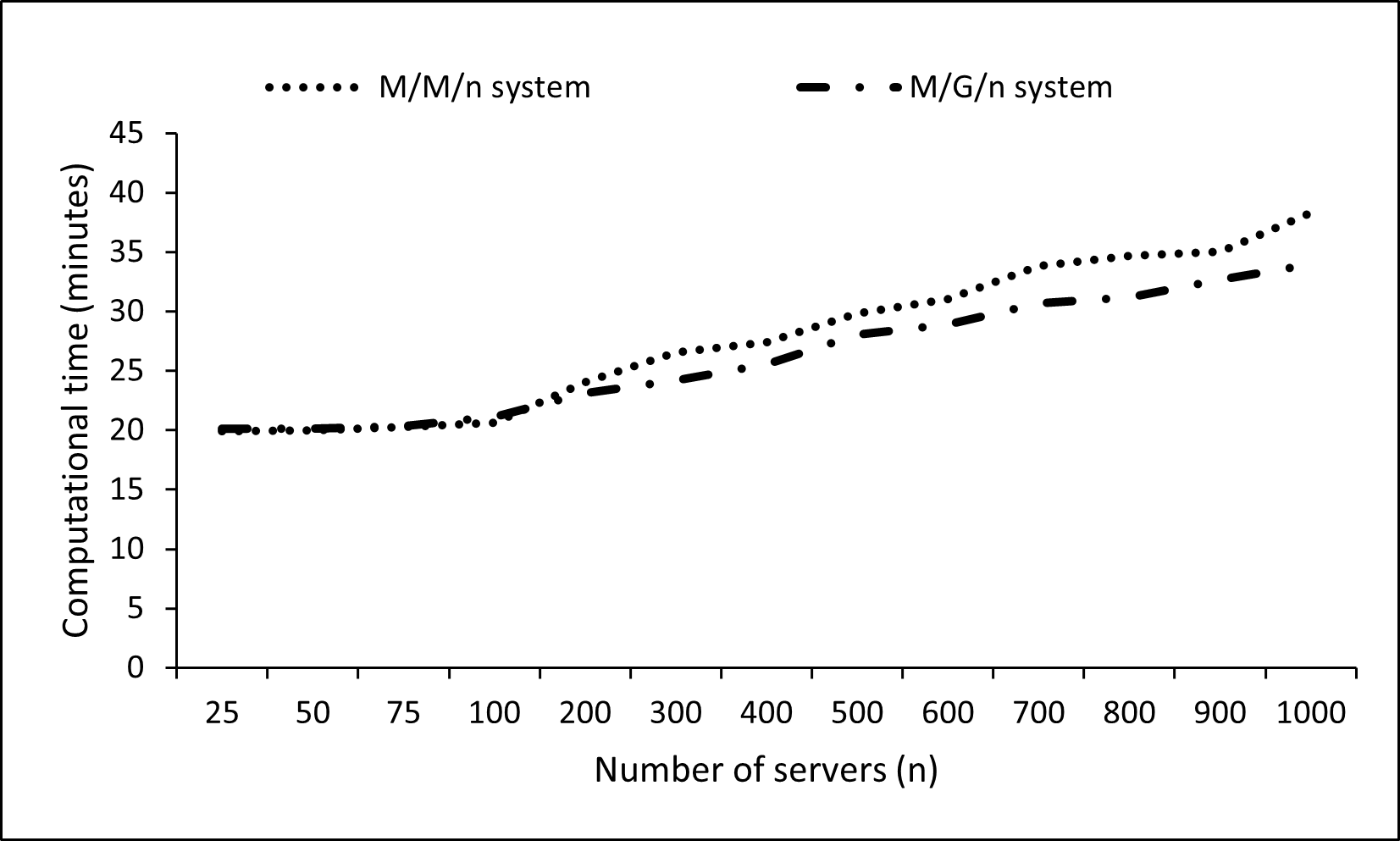}
    \caption{The effect of the number of servers on the computational runtime for simulations of $M/M/n$ and $M/G/n$ systems.}
    \label{scaling}
\end{figure}

Additionally, we investigated how the change in the number of servers within a subsystem affects simulation runtime. To do so, we conducted a series of experiments using single-stage $M/G/n$ and $M/M/n$ systems. The number of servers ($n$) was varied between 25 and 1000 and computational runtimes were recorded at each level. The server occupancy level was kept constant for these experiments by adjusting the service rates of each server in accordance with the number of servers at the system in question. The outcomes recorded are the average runtimes from 30 independent replications, and are presented via Figure~\ref{scaling}. The results demonstrate that the number of servers had no visible impact on the computational runtime when the server count was less than 100 servers, beyond which an approximately linear increase in the runtime can be observed. A potential explanation for this observation involves the notion that at moderate or high workstation occupancy levels (e.g., $>$ 50\%), searching for the first idle server requires more simulation runtime when the number of servers becomes large.  Hence for this study, we assume that the number of servers within each subsystem is significantly less than 100, and as such, we exclude the number of servers as a contributing factor to RS. This experiment and the attendant observations also motivated our exclusion of one form of the `aggregation' simplification operation that involves replacing multiple servers in a workstation / subsystem with a single server with an appropriately scaled service rate. However, our proposed RS prediction framework is applicable to the practice of aggregating multiple subsystems into a single subsystem, which is another (more widely practised) form of the aggregation simplification operation. This will be discussed in more detail once the RS prediction methodology itself is described.

From this discussion, the following primary principle underpinning our RS prediction framework can be derived. 

The RS achieved with a simplified model relative to its parent model is primarily caused by the reduction in the total number of instructions executed. Further, the total RIE itself depends on the server occupancy indicating that characterising the relationships between the server occupancy and RIE, and then between RIE and the RS attained by abstraction, will comprise the RS prediction framework. 

We now describe the process of characterising these relationships.

\subsection{Runtime Savings Prediction Model Development}
\label{sec:crr-model-development}

At this stage, we reiterate that, once the RS prediction models are available, the RS prediction is made before the simplified model is developed, and ideally without running the parent simulation even once for the purposes of model simplification. However, per the discussion in the previous subsection, the RS prediction framework will require at minimum occupancy levels of the subsystems being abstracted out, which implies that the parent simulation must have been executed at least once (for analyses of general interest, and not for the purpose of model simplification) from which key simulation outcomes are available. These would include the subsystem occupancy values and, although this is not strictly required (see Section~\ref{sec:implem} for a more detailed discussion regarding this), the numbers of entity arrivals to the system over simulation run lengths of interest. 

The RS prediction framework is developed in two stages. In the first stage, a model that relates server occupancy and the average number of instructions per arrival is derived for both the parent model and the simplified model. Outputs from these models are combined with the average number of arrivals to the subsystem subject to simplification to obtain the associated RIE. In the second stage, a model that relates the RS and RIE is developed.

In order to develop the above models, we considered simulations of simple queueing systems, for which we present the notation below.

\begin{itemize}
    \item `\textit{2s}' denotes a two-stage tandem queueing system simulation shown in Figure~\ref{figure:abstraction}a. The system is composed of a generator and two identical single-server subsystems in series.
    \item `\textit{1s}' refers to a single-stage queueing system DES with a generator, a queue, and a server. 
    \item `\textit{ss}' is used to refer to a single (sub)system DES without a generator. It is essentially the second-stage system from a \textit{2s} system.
    \item `\textit{ms}' denotes the notation for the simplified version of the \textit{2s} system. It is constructed by substituting the second stage (\textit{ss} stage) with a random variable modelling length-of-stay in that subsystem and can be seen in Figure~\ref{figure:abstraction}b. 
    
\end{itemize}

At this point, we pause to emphasize that as part of the simplification operation, the subsystems that are abstracted out are typically \textit{ss} systems, and consequently the RIE from these operations can be estimated as a sum of the number of instructions associated with each \textit{ss} system abstracted out. Here, the RIE itself will be calculated using as a function of the server occupancy levels of the subsystem(s) in question. Algorithm~\ref{algo:absrspred} in Section~\ref{sec:implem} provides a formal description of this process. This is the rationale for developing a model for the relationship between the server occupancy and number of instructions per arrival of an \textit{ss} system. However, unlike the \textit{2s, 1s} and \textit{ms} systems, the average number of instructions per arrival associated with an \textit{ss} system cannot be estimated directly via a DES given that a component generator for creating the entities will be required for any queueing system simulation. Hence the analysis for the \textit{ss} system was done by subtracting the average number of instructions per arrival for the \textit{1s} system from that for the \textit{2s} system.

We now describe both stages of the RS prediction model development process. 

\newlist{subroutine}{enumerate}{1}
\setlist[subroutine,1]{label=(\alph*)}
\par\noindent\rule{\textwidth}{0.4pt}
\noindent \textbf{Stage 1.} Model derived: Average number of instructions per arrival versus server occupancy. Models are derived for the $k^{th}$ system, where $k \in K = \{2s, 1s, ms\}$, and with the $m^{th}$ queueing discipline, where $m  \in M = \{ M/M/1, M/G/1, G/G/1 \}$.  $n$ server occupancy levels are used to fit the model, where occupancy levels $\rho_i$ take values in $P = \{\rho_1,\rho_2, ...\rho_n\}$.  \\
 
\noindent I. For the $k^{th}~(k \in K)$ system with the $m^{th}$ subsystem configuration, $ m \in M$, the following steps are carried out for each server occupancy value $\rho_i \in P$.

\begin{subroutine}
   \item  Run the DES and generate the trace.  
   \item Utilise the trace for the $i^{th}$ server occupancy level to calculate and record the average number of instructions per arrival, denoted as $\theta^{m}_{k}(\rho_i)$.
\end{subroutine}

\noindent II. Based on the data collected from step 1.I(a), for each $m \in M$ and $k \in K$, develop a regression model, $\theta^{m}_k = f(\rho |\beta^m_k, \epsilon^m_k)$, where $\beta^m_k$ parameterises the model and $\epsilon^m_k$ captures the stochasticity in the relationship.
\vspace{0.5 cm}
 
\noindent \textbf{Stage 2.} Model derived: RS as a function of the RIE achieved with the simplified model.\\

\noindent I.  Carry out the following steps for $m \in M$ and $k \in \{2s,ms\} $ systems, for each server occupancy value $\rho_i \in P$:
  \begin{subroutine}
  \item From the simulation executions in Step 1.I(a), record the average execution time ($t^{m,k}_{i}$) and the average number of arrivals ($a^{m,k}_{i}$) for each of the $\rho_i$ server occupancy values. 
  \item Using the regression model derived in Stage I, estimate the expected number of instructions per arrival as a function of server occupancy $\rho_i$ ($\theta^m_{k}(\rho_i)$).
  \item  Estimate the average total number of instructions across the simulation run length of interest for the $i^{th}$ server occupancy level as $I^{m,k}_{i} = a^{m,k}_{i} \times \theta^m_{k,i}$.
  \end{subroutine}
  
\noindent II. For each value of $I^{m,k}_i$ and $t^{m,k}_i$ $(k \in \{ 2s, ms\})$ collected in 2.I, perform the following steps.

  \begin{subroutine}
  \item Compute the difference between $I^{m,2s}_i$ \& $I^{m,ms}_i$ to obtain the RIE as: $\Bar{I}^m_i = I^{m,2s}_{i} - I^{m,ms}_{i}$.
  \item Similarly, calculate the average RS as: $\phi^m_i = t^{m,2s}_i - t^{m, ms}_i$.
  \end{subroutine}

\noindent III. Using the data collected in 2.II, fit a regression model $\phi^m = g(\bar{I}^m|\alpha^m, \eta^m)$, where $\alpha^m$ and $\eta^m$ parameterise the model.
\par\noindent\rule{\textwidth}{0.4pt}


The computational implementation of the RS prediction model development process for an example case is demonstrated in Algorithms \ref{algo:inst} and~\ref{algo:savpred}. We then move to Section~\ref{sec:implem} for the description of the numerical experiments conducted to implement this RS prediction model development process.

\begin{algorithm}[htbp]
    \caption{Pseudocode for the implementation of Stage 1 of the RS prediction model development process.}
    \label{algo:inst}
    \begin{algorithmic}[1]
        \Function{simulate}{$\lambda$}
            
            \State Initialise simulation time \& warm-up time 
            \State Simulate the model for arrival rate $\lambda$
            \State Record simulation trace in an array (TraceList)
            \State \Return TraceList
        \EndFunction

        \Function{countInstructions}{TraceList, $n$}
            \State Read TraceList
            \State Remove first and last ten entries from the trace
            \State Sample n ($=100$) values at random from TraceList, sampled\_trace $\gets$ TraceList 
            \State Compute average number of instructions per arrival, $\theta$ = average(sampled\_trace)
            \State \Return $\theta$
        \EndFunction
        
        \Comment{Begin Procedure}
        
        \State Specify: 
            Server occupancy values $\rho \in P = \{\rho_1, \rho_2, ...,\rho_n\}$, mean service rate ($\mu$) and the number of replications ($r$) to be performed for each occupancy value
        \State Initialise an array (NumInstructionsList), for data collection
        
        \For{$\rho \in P $}
            \State Initialise an array for data collection, AverageList
            \State $j \gets r$
            \State Calculate arrival rate, $\lambda = \rho \times \mu$
            \While{$j \neq 0 $}
                \State Execute simulation \& obtain trace as output in an array, OutputList $\gets$ \Call {Simulate}{$\lambda$}
                \State Estimate average instructions per arrival, $\theta_{\rho,r} \gets$ \Call{countInstructions}{OutputList, 100}
                    
                \State Store $\theta_{\rho,r}$ in an array(AverageList), addItem(AverageList, $\theta_{\rho,r}$)
                \State $j = j-1$
            \EndWhile
            \State Calculate average value of $\theta_{\rho}$ over $r$ replicates \& store the output, addItem(NumInstructionsList, $\bar{\theta_{\rho}}$ = average(AverageList))            
        \EndFor
        
        \Comment{End Procedure}
    \State Regress the average number of instructions per arrival on server occupancy to characterise $\theta = f(\rho | \beta, \epsilon)$ \;

    \end{algorithmic}
\end{algorithm}

\begin{algorithm}[htbp]
    \caption{Pseudocode for the implementation of Stage 2 of the RS prediction model development process.}
    \label{algo:savpred}
    \begin{algorithmic}[1]
        \State Specify:  Server occupancy values $\rho \in P = \{\rho_1, \rho_2, ...,\rho_n\}$, mean service rate ($\mu$), \& number of replications ($r$) to be performed for each $\rho$
            
        \State Initialise arrays for data collection; ArrNumOfInstruction, ArrRunTime  
        \State Compute average arrival rates $A = \{\lambda : \lambda = \rho \times \mu,~ \rho \in P\}$
        
        \For{$k \in \{2s, ms\}$}
            \For{$\rho \in P$}
                \State Initialise arrays RepTimeList() \& RepArrivalList() for data collection                
                \State $i \gets r$
                \While{$i \neq 0$}
                    \State Execute simulation for $\lambda$ and $\rho$ values
                    \State Collect number of arrivals ($n$) \& model run time ($t$) 
                    \State Update arrays, addItem(RepArrivalList, $n$)  \& addItem(RepTimeList, $t$) 
                    \State Update $i =i-1$
                    \EndWhile
                \State Compute average number of arrivals \& runtime over $r$ as;
                \Indent
                \State $t^k_{\rho}$ = average(RepTimeList) \& $n^k_{\rho}$ = average(RepArrivalList)
                \EndIndent
                \State Update arrays:
                addItem(ArrTime, $t^k_{\rho}$)
                addItem(ArrInstructs, $n^k_{\rho}$)
            \EndFor
            \State Estimate $\bar{\theta}^{k}_{\rho}$ as a function of $\rho$ from Stage 1
            \State Compute $I^{k}_{\rho} = \bar{\theta}^{k}_{\rho} \times n^k_{\rho}$  
            \State Record average run time and number of instructions as
            \Indent
                \State addItem(ArrNumOfInstructions, $I^{k}_{\rho}$), addItem(ArrTime, $t^{k}_{\rho}$)
            \EndIndent
        \EndFor
        \State Calculate the difference between the number of instructions, $\bar{I}$ = ArrNumOfInstructions$[2s]$   - ArrNumOfInstructions$[ms]$ 
        \State Calculate RS, $\phi = $ ArrTime$[2s]$   - ArrTime$[ms]$ 
        \State Regress $\phi$ on $\bar{I}$
    \end{algorithmic}
\end{algorithm}

\normalsize

\section{Runtime Savings Prediction Model Development: Computational Implementation}
\label{sec:implem}
In this section, we describe the computational experiments carried out to develop the RS prediction models. All computation was performed on a Windows 11 64-bit workstation, equipped with an Intel $i7$ 4-core processor with a base clock speed of $3.3$ gigaHertz and $16$ gigabytes of memory. 


To develop models relating server occupancy with the average number of instructions per arrival $\theta$, we constructed DESs of the $2s$, $ms$, and $1s$ systems. The service and arrival patterns within these systems were then adjusted to represent the $M/M/$, $M/G/$, and $G/G/$ queueing disciplines. 

We first developed a DES model of the $2s$ system (the parent model). This parent model was then modified to represent a) the single-stage ($1s$) system and b) the $ms$ system. The $1s$ system was constructed by omitting the second stage from the $2s$ model. The following procedure is employed for the development of the $ms$ DES system. In the $ms$ system, the full-featured second stage subsystem from the parent ($2s$) model is replaced by a random variable that represents the LOS within that stage. The parameterisation of the random variable is accomplished by collecting the LOS data from the second stage subsystem of the parent model and performing a curve-fitting procedure. However, if standard parametric distributions do not provide an acceptable level of fit, a nonparametric approach, specifically kernel density estimation (KDE), is employed to find the second stage subsystem LOS distribution. The second stage subsystem in the parent model is then replaced with its corresponding LOS random variable. Subsequently, the simplified model is executed and data for the overall LOS in the system is recorded. Next, the overall system LOS data from the simplified model and the parent models are used in a Kullback-Leibler (KL) divergence test procedure to compare their distributions. 
This procedure attempts to preserve the stochastic character of the simulation outputs of interest. 

We demonstrate this model simplification procedure with three examples, one from each queueing system type ($M/M/, M/G/$, and $G/G/$), and with one randomly selected server occupancy value for each system type. Detailed information regarding the second-stage subsystem LOS modelling and the validation processes are provided in Table \ref{tab:abs1}. For the $M/M/$ case, first the $2s$ simulation model was executed and LOS values were recorded for both the entire system and the second stage. Then, as outlined above, this dataset underwent a curve-fitting procedure using the \textit{stats} module of Python's \textit{SciPy} \citep{scipy} library and the best fitting distribution was selected based on the $p$-value of the Kolmogorov-Smirnov (KS) test. The simplified model was constructed using the LOS random variable for the second stage subsystem and subsequently executed. The overall system LOS data generated by the simplified and parent models were compared via the KL divergence test. A KL divergence value of 0.04 was obtained indicating that the overall system LOS distributions are statistically similar. The same procedure was followed for the $G/G/$ and $M/G/$ systems; however, for the $M/G/$ case, none of the standard parametric distributions available in the \textit{SciPy} library provided an acceptable level of fit ($p$-value $> 0.05$) to the second stage subsystem LOS data. Consequently, the LOS distribution was estimated via kernel density estimation implemented using the Scikit-learn library \citep{pedregosa2011scikit}. The KDE estimator of the second stage subsystem LOS distribution was then used to construct the $ms$ model. Histograms comparing the LOS data from the parent and child models can be found in \ref{app:asbtraction}. The codes used for the curve-fitting process, KDE and validation can be accessed from \ref{app:github}. Furthermore, the model execution time of the simplified model remains unaffected by whether a parametric distribution or a KDE estimator was used to model the second-stage subsystem LOS. Overall, from Table~\ref{tab:abs1}, it is evident that the abstraction operation yields simulation outcomes comparable to that of the parent simulation model.

\begin{table}[htbp]
    \centering
    \caption{Example length of stay distribution modelling and validation results for the abstraction model simplification operation applied to three queueing systems.}
    \small
    \begin{tabular}{|p {1.25 cm}|p {1.7 cm}|p {7.3 cm}|p {2.5 cm}|}
    \hline
    Queueing System & Occupancy  & Abstracted subsystem length of stay modelling & KL Divergence Estimate \\ \hline
    $M/M/$ & $68\% $  & Method: parametric; Distribution: Beta prime ($1.03, 1.37\times10^6, -8.58\times10^6, 5.01\times10^6 $); $p$-value: $ 0.58$  & $0.04$\\ \hline
    $M/G/$ & $75\% $  & Method: non-parametric (KDE); 
    Kernel: Gaussian; Bandwidth: 0.1  & $0.14$ \\ \hline
    $G/G/$ & $71\% $  & Method: parametric; Distribution: Generalised Logistic ($14.76, 0.74, 1.54$); $p$-value: $0.99$ & $0.09$\\ \hline
    \end{tabular}
    
    \label{tab:abs1}
\end{table}

We now discuss the implementation of the RS prediction model development process discussed in Section~\ref{sec:method}. We begin by providing information regarding the parameterisation of the subsystems within the $ M/M/$, $M/G/$ and $ G/G/$ systems in Table \ref{tab:input_para}. In the $M/M/$ parent DES, interarrival times and service times at each server followed independent exponential distributions. The servers in both the first and second stages had a mean service time of 1 minute. For the $G/G/$ case, both subsystems service rates were modelled using a Gamma distribution, maintaining equal mean service times of 5 minutes with a standard deviation of $1.5$ minutes. Furthermore, interarrival times were governed by a Gaussian distribution. In the $M/G/$ case, both the interarrival and the first subsystem's service times were modelled using exponential distributions, while the second subsystem’s service time was modelled with a Gaussian distribution. A lower bound of 0.01 was imposed on the Gaussian distribution to avoid negative values.

\begin{table}[!ht]
    \centering
    \caption{Parameterisation of the two stage and single stage queueing system simulations used for developing the runtime savings prediction models.}
    \small
    \label{tab:input_para}
    \begin{tabular}{|C{2cm}|p{3.5cm}|p{4cm}|p{4cm}|}
    \hline
    \multirow{2}{*}{System}  & \multirow{2}{*}{Arrival} &  \multicolumn{2}{|c|}{Service} \\ \cline{3-4} 
         
           &        & Subsystem 1 & Subsystem 2 \\ \hline
        
    Two stage $M/M/$ & Exponential ($\lambda$) & Exponential (1) & Exponential (1) \\ \hline
        
    Single stage $M/M/$ & Exponential ($\lambda$) & Exponential (1) & NA \\ \hline
        
    Two stage $M/G/$ & Exponential ($\lambda$) & Exponential (2) & Normal (2, 0.2); bounded at 0.01 \\ \hline
    Single stage $M/G/$ & Exponential ($\lambda$) & Exponential (2) & NA \\ \hline
        Two stage $G/G/$ & Normal ($\lambda$, 1.5); bounded at 0.01 & Gamma (11.11, 0.45); $\mu = 5$, $\sigma = 1.5$ & Gamma (11.11, 0.45); $\mu = 5$, $\sigma = 1.5$ \\ \hline
        Single stage $G/G/$ & Normal ($\lambda$, 1.5); bounded at 0.01 & Gamma (11.11, 0.45); $\mu = 5$, $\sigma = 1.5$ & NA \\ \hline
    \end{tabular}
    \textit{Note: $\lambda = \frac{\mu}{\rho}$, $\mu =$ mean service time, $\sigma = $ standard deviation of service time, $\rho=$ server occupancy, and $ \rho \in $ \{$0.2, 0.21, ...0.93$\}; NA = not applicable.}
\end{table}

\begin{figure}[!htbp]
    \centering
    \includegraphics[width=0.6\linewidth]{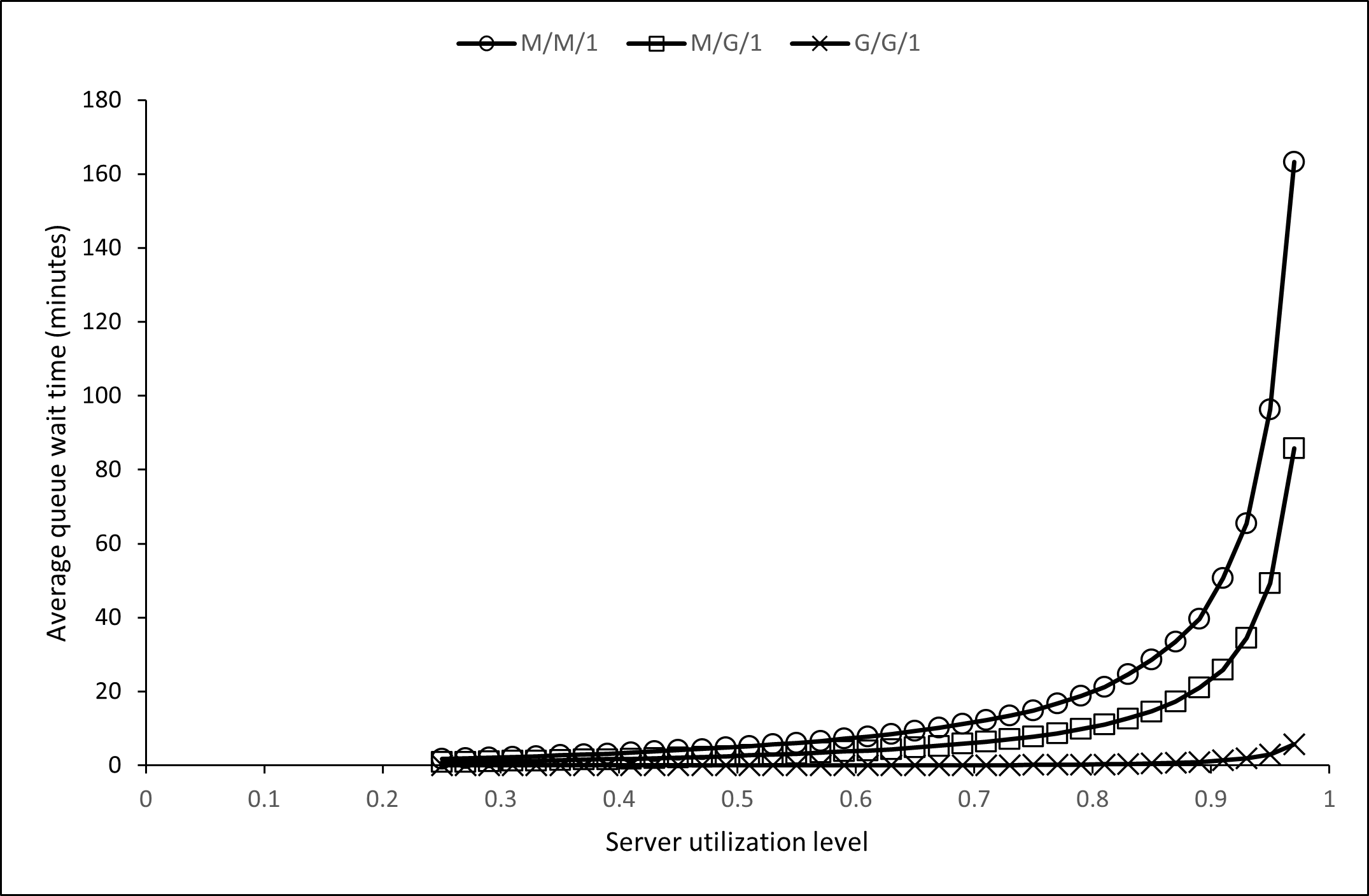}
    \caption{Average waiting time versus server occupancy for $M/M/1$, $M/G/1$, and $G/G/1$ queueing systems.}
    \label{fig:1}
\end{figure}

\underline{Stage 1 demonstration.} We recall that in Stage 1, models that estimate $\theta$ as a function of server occupancy are developed. 
The models were initialised with a warm-up time of $200$ days of simulation time after which they were allowed to run for $10$ days of simulation time.
To collect the required data, models were executed for server occupancy levels ranging from 20\% to 93\% with a step size of 1\%. This range was adopted because of the following two reasons. Firstly, in most physical systems, server occupancy in the range of $80\%$ to $90\%$ is considered desirable~\citep{Agarwal}. 
Secondly, beyond this range the average queue waiting time rises very steeply, which is shown in Figure \ref{fig:1}. 
Next, for each occupancy value, $\theta$ was recorded. This was achieved by following the procedure outlined in Algorithm \ref{algo:inst}.

The parent model was executed (10 replications) at predefined server occupancy values. For every replication, trace output was recorded and then analysed using a custom Python function we developed to quantify the number of instructions for each entry (arrival/entity) listed within the trace. Next, from the output of the previous step, one hundred entries were randomly selected, and their respective number of instructions was averaged to compute the average number of instructions per arrival. This average was further aggregated across the ten replications to derive $\theta$ for a given server occupancy value. Following this, the data recorded in the above manner for the server occupancy range underwent a regression model fitting procedure with server occupancy serving as the independent variable and $\theta$ as the dependent variable. 

\begin{figure}[htbp]
  \centering
  \begin{subfigure}[htbp]{0.55\linewidth}
    \includegraphics[width=1.1\linewidth]{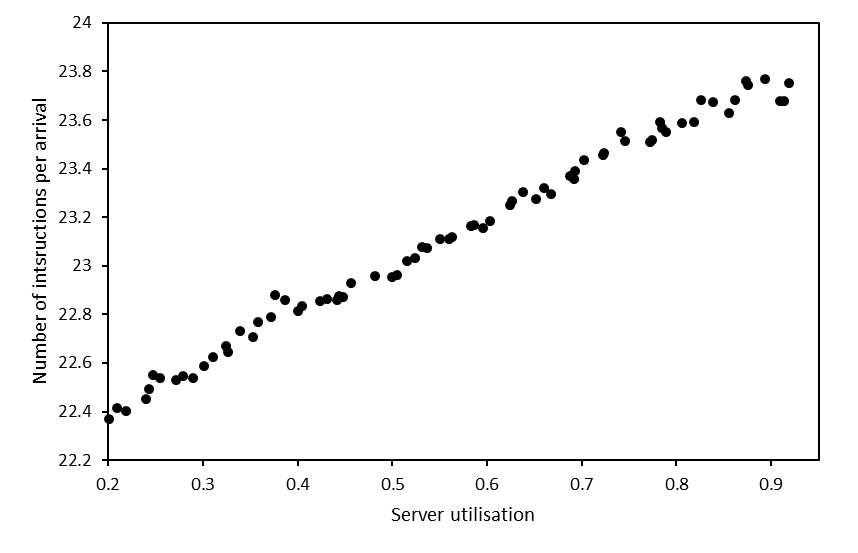}
    \caption{$M/M/$ system.}
    \label{fig:sub0}
  \end{subfigure}
  \hfill 
  
  \begin{subfigure}[htbp]{0.48\linewidth}
    \includegraphics[width=1.1\linewidth]{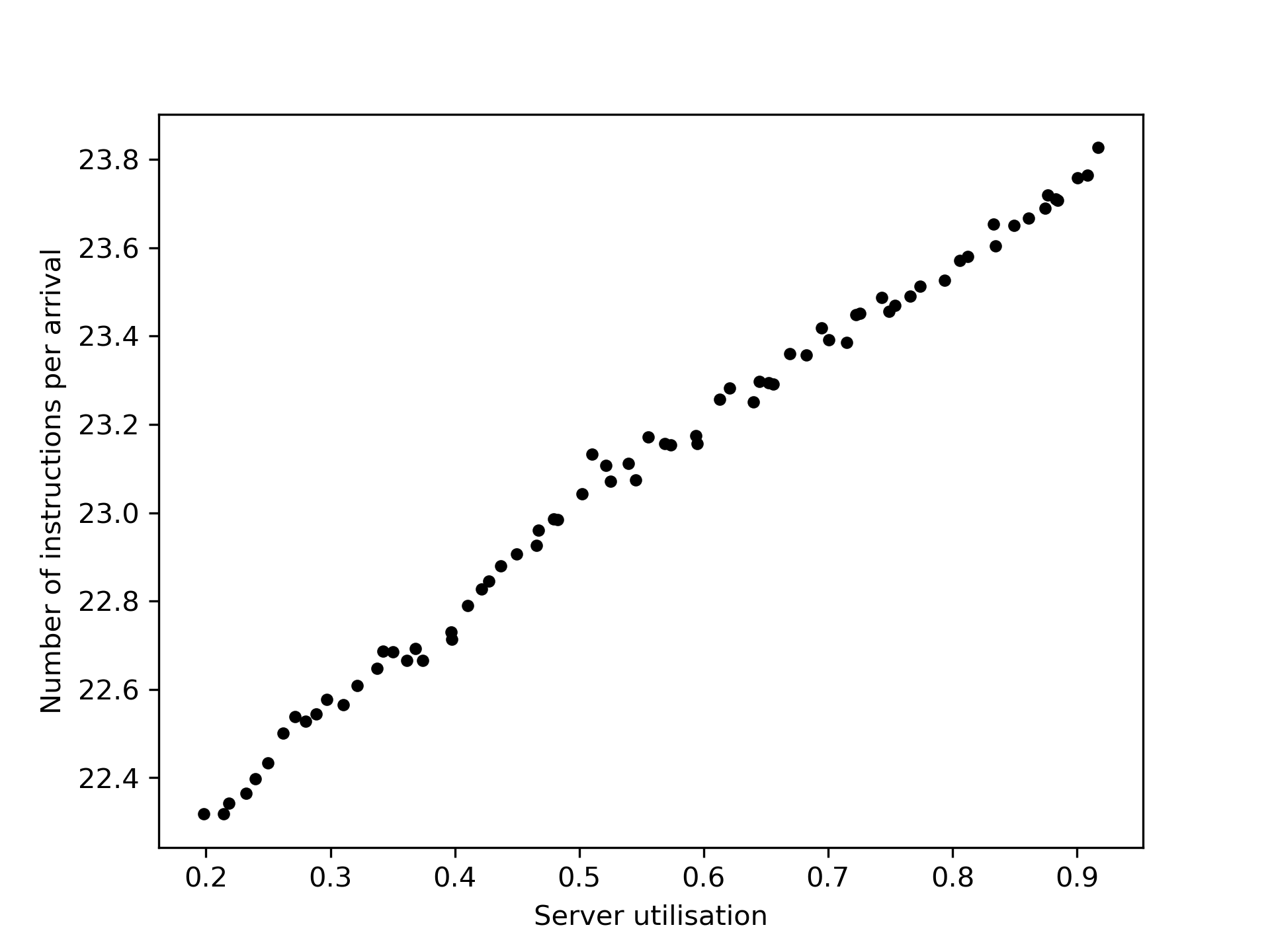}
    \caption{$M/G/$ system.}
    \label{fig:sub1}
  \end{subfigure}
  \hfill 
  \begin{subfigure}[htbp]{0.48\linewidth}
    \includegraphics[width=1.1\linewidth]{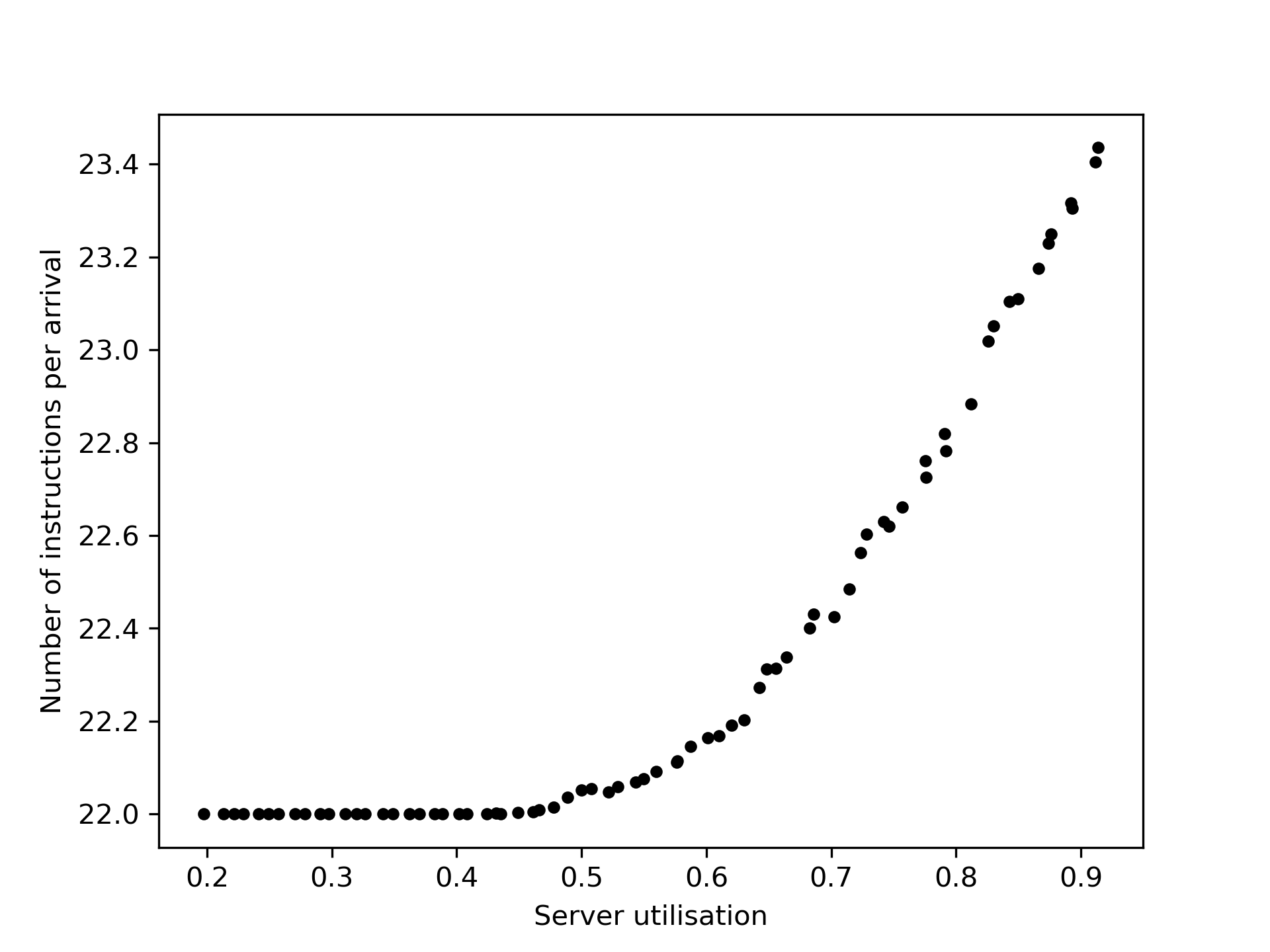}
    \caption{$G/G/$ system.}
    \label{fig:sub2}
  \end{subfigure}
  \caption{The relationship between the average number of instructions per arrival and server occupancy for the second subsystem within two-stage systems with $M/M/$, $M/G/$ and $G/G/$ systems.}
  \label{fig:main}
\end{figure}

In Figure~\ref{fig:main}, we present visualisations of the relationship between $\theta$ and second-stage subsystem server occupancy for the \textit{2s} $M/M/1$, $M/G/1$ and $G/G/1$ systems. A linear correlation between the two variables is evident in the $M/M/1$ case, as depicted in Figure~\ref{fig:sub0}. Further, from Figure~\ref{fig:sub1}, the relationship between the two variables in the $M/G/1$ case appears to be nonlinear, displaying a concave shape.
Conversely, Figure~\ref{fig:sub2} reveals an interesting pattern for the $G/G/1$ case where $\theta$ remains relatively constant from server occupancies between 20\% and about 50\%. A similar trend was also observed in the single-stage $G/G/1$ system. Based on these observations, we truncated the simulation data to exclude values up to 50\% server occupancy.  
The results of the regression, including the $R$-squared values, are presented in Table \ref{tab:crr-equations}.

Importantly, we note here that for the simplified ($ms$) system we found the second stage or the LOS random variable accounted for a fixed two instructions per arrival. This is due to the fact that there are no simulation components in this stage, such as queues and servers, and hence the processor executed only two instructions for each arriving entity to hold it in the system for the sampled length of time and then to release the entity. Consequently, the simplified \textit{ms} system could be represented as a single-server ($1s$) system with only two additional instructions per arrival, i.e, $\theta_{ms} = \theta_{1s} + 2$.

\underline{Stage 2 demonstration.} The second stage of the process involves developing models that take the reduction in the average number of instructions executed (RIE) as input to generate expected RS as output. This was accomplished by following the steps listed in Algorithm \ref{algo:savpred}.
The stage 2 process required executing both the $2s$ DES and its \textit{ms} DES, while tracking the number of arrivals (notated as $n$) and the computational runtime for each model. These simulations were performed for the same server occupancy range as above - i.e., between 20\% and 93\%, incremented in steps of 1\%. 
Utilising the equations for $\theta_{2s}$ and $\theta_{ms}$ $ (=\theta_{1s} + 2)$ derived in stage 1, $\theta$ values for both the models at each occupancy value were computed. Next, the total instructions were determined by multiplying the $\theta$ with total arrivals ($n$). Subsequent to this, the RIE and RS achieved with the simplified model were calculated using data from the above steps. Then, a regression model was fitted to the data using RIE and RS as the independent and dependent variables, respectively. 

The resulting RS versus RIE regression equations along with their respective $R$-squared values are displayed along with the $\theta$ versus server occupancy regression equations in Table~\ref{tab:crr-equations}.
We also checked whether the observed pattern for the $G/G/$ case is specific to the choice of arrival and service pattern distributions. To do so, we experimented with different combinations of arrival and service rate distributions that included Gaussian, uniform, triangular, and beta, among others. Our experiments revealed similar relationships between $\theta$ and server occupancy.

\begin{table}[!htbp]
\centering
\caption{Regression equations part of the runtime savings prediction framework for model simplification operations.}
\small
\renewcommand{\arraystretch}{1.2}
\begin{tabular}{|l|l|l|l|}
\hline
DES           & Queueing system model        & Regression equation & R-squared value \\ \hline
\multirow{4}{*}{$M/M/$} & Two-stage    & $\theta^{mm}_{2s} = 1.97x + 33.02$          & 0.99         \\ \cline{2-4} 
                        
                     & Single-stage    & $\theta^{mm}_{1s} = 0.99 x + 13.00$          & 0.99         \\ \cline{2-4} 
                     & Single-server  & $\theta^{mm}_{ss} = 0.97 x +9.02 $          & 0.92         \\ \cline{2-4} 
                     & RS prediction &    $\phi_{mm} = 0.04x' - 0.05$       & 0.99         \\ \hline
\multirow{4}{*}{$M/G/$} & Two-stage      & $\theta^{mg}_{2s}$ = $-0.69 x^2 + 2.80 x + 21.78 $          & 0.99        \\ \cline{2-4} 
                     & Single-stage    & $\theta^{mg}_{1s}$ = $1.03 x + 12.97 x$           & 0.99         \\ \cline{2-4} 
                     & Single-server  & $\theta^{mg}_{ss}$ = $-0.47 x^2 + 1.51x +9.87$         & 0.98         \\ \cline{2-4} 
                     & RS prediction & $\phi_{mg} = 2.28 \times 10^{-5} x'^2 + 0.03x'+ 0.091$          & 0.99         \\ \hline
\multirow{4}{*}{$G/G/$} & Two-stage      & $\theta^{gg}_{2s}$ = $5.90 x^2 - 4.85 x + 22.96 $          & 0.99         \\ \cline{2-4} 
                     & Single-stage    & $\theta^{gg}_{1s}$ = $3.61 x^2 -3.40 x + 13.79 x$          & 0.99         \\ \cline{2-4} 
                     & Single-server  & $\theta^{gg}_{ss}$ = $2.48 x^2 - 1.76 x + 9.29$          & 0.98         \\ \cline{2-4} 
                     & RS prediction & $\phi_{gg} = 0.03 x + 0.07$          & 0.99         \\ \hline
\end{tabular}

\label{tab:crr-equations}

\footnotesize
    \textit{Note: $x$ stands for server occupancy. $x'$ is the reduction in the number of instructions between the parent and simplified models in $10000$s, and $\phi$ is the expected runtime savings. For all the cases, $\theta_{ms} = \theta_{1s}+2$}.
\end{table}

The calculation of the total number of instructions was based on the average total number of arrivals ($n$), which we collected directly from the simulation outputs. An alternate approach to calculate this value could involve multiplying the average arrival rate, derived from the server occupancy and the mean service rate, by the run length of the simulation experiment. In constructing the RS prediction model, we also tested this estimation technique for $n$ and found that it provided results that were comparable.

We conclude this section by describing the method for applying the RS prediction framework to a potential model simplification operation. Consider a DES of a system with $m$ subsystems, each with recorded occupancy $\rho_i~(i\in M =\{1,2,\dots,m\}$, where $M$ is the index set of subsystems comprising the DES. In order to describe the method for deploying the RS prediction framework, we need to formally specify the types of simplification operations for which our framework can be used: (a) $k_1$ subsystems ($k_1 < m$) are abstracted out with $k_1$ LOS random variables (denoted by $S1$); (b) $k_2~ (< m)$ subsystems are abstracted out and replaced by a single random variable (denoted by $S2$); (c) $k_3~ (< m)$ subsystems are replaced by a single subsystem (denoted by $S3$) (an aggregation operation); (d) a combination of more than one of the above three types - for example, a simplification operation wherein $k_1$ subsystems undergo the $S1$ operation, $k_2$ subsystems undergo the $S2$ operation, and $k_3$ subsystems undergo the $S3$ operation, with the condition that $k_1 + k_2 + k_3 < m$. Note that if $k_1 + k_2 + k_3 = m$, we are no longer concerned with model simplification, but enter the realm of simulation metamodelling. In general, for the purposes of our study, we can specify a model specification operation as a tuple $O = (K_1, K_2, K_3)$, where each $K_i \subset M$ (with $|K_i| = k_i~(i\in \{1,2,3\}$ and $k_1 +k_2 + k_3 < m$) indexes the subsystems subject to each of the three operations.

We provide in Algorithm~\ref{algo:absrspred} below a description of how to estimate RS for a model simplification operation using our proposed predictive framework. 

\begin{algorithm}[htbp]
    \caption{Pseudocode for the deployment of the RS prediction framework for model simplification operations.}
    \label{algo:absrspred}
    \begin{algorithmic}[1]
        \State Initialise with information regarding parent DES: $m$ workstations/subsystems with workstation index set $M = \{1,2,\dots, m\}$
        \State Initialise with specification of model simplification operation $O = (K_1, K_2, K_3)$ 
        \State Initialise with information from parent DES: server occupancy vector for the parent DES $\rho = \{\rho_1, \rho_2,\dots,\rho_m\}$
        \State Initialise with information from parent DES: number of arrivals to the sets of subsystems corresponding to each element of $O$: $n_{Sl} = \{n_s: s \in K_l\}$, with $n_{sim2}$ and $n_{sim3}$ representing the overall number of arrivals to the sets of subsystems subject to simplification operation $S2$ and $S3$
        
        \State Initialise with regression equations for the average number of instructions per arrival $\theta_{ss}^v$ and expected RS $\phi_v$ from the RS prediction framework ($v \in \{mm,mg,gg\}$)
        \State Initialise storage variables $\bar{I}_1, \bar{I}_2, \bar{I}_3 = 0$
        \For{$j \in \{1,2,3\}$}
        \For{$i \in K_j$}
        \State Identify queueing discipline $v \in \{mm,mg,gg\}$ of the $i^{th}$ subsystem and relevant equations $\theta^{v}_{ss}$ and $\phi_v$ from RS prediction framework
        \State Estimate $\theta^v_{ss,i} = f^{v}_{ss}(\rho_i)$
        \State Estimate total number of instructions associated with subsystem $I_{j,i} = \theta^v_{ss,i} \times n_i$ ($n_i \in n_{Sj}$) 
        \State Estimate RIE w.r.t. removing $i^{th}$ subsystem $\bar{I}_j = \bar{I}_j + I_{j,i}$
        \EndFor
        \State If $j$ = 1:
            \Indent \Indent
            \State $\bar{I}_1 = \bar{I}_1 - 2\times \sum\limits_{l \in n_{S1}} n_l$
            \State Estimate expected RS $\phi_{1} = g_v(\bar{I}_1)$
            \EndIndent \EndIndent
        \State Else If $j$ = 2:
            \Indent \Indent
            \State $\bar{I}_2 = \bar{I}_2 - 2\times n_{sim2}$
            \State Estimate expected RS $\phi_{2} = g_v(\bar{I}_2)$
            \EndIndent \EndIndent
        \State Else:
            \Indent \Indent
            \State $\rho_{sim}$ = server occupancy of subsystem replacing subsystems $\in K_3$ 
            \State Identify queueing discipline $v \in \{mm,mg,gg\}$ of the simplifying subsystem and relevant equations $\theta^{v}_{ss}$ and $\phi_v$
            \State Estimate $\theta^v_{ss,sim} = f^{v}_{ss}(\rho_{sim})$
            \State Estimate total number of instructions associated with subsystem $I_{sim} = \theta^v_{ss,sim} \times n_{sim3}$
            \State Estimate RIE as $\bar{I}_3 = \bar{I}_3 - I_{sim}$
            \State Estimate expected RS $\phi_{3} = g_v(\bar{I}_3)$
            \EndIndent \EndIndent
        \EndFor
        \State Return expected RS estimate as $\phi = \phi_1+\phi_2+\phi_3$
    \end{algorithmic}
\end{algorithm}

Algorithm~\ref{algo:absrspred} makes it clear as to why incorporating the $ss$ system in the RS prediction framework is important - while the $2s$ and $ms$ systems are used to generate the equations for RS prediction, the subsystems subject to simplification can for the most part be modelled as $ss$ subsystems. Further, it can also be seen how, even though the models within the RS prediction framework were characterised only using the abstraction operation, they can be applied to aggregation operations as well.

\section{Runtime Savings Prediction Model: Deployment and Validation}
\label{sec:valid}

\begin{figure}
    \centering
    \includegraphics[width=\linewidth]{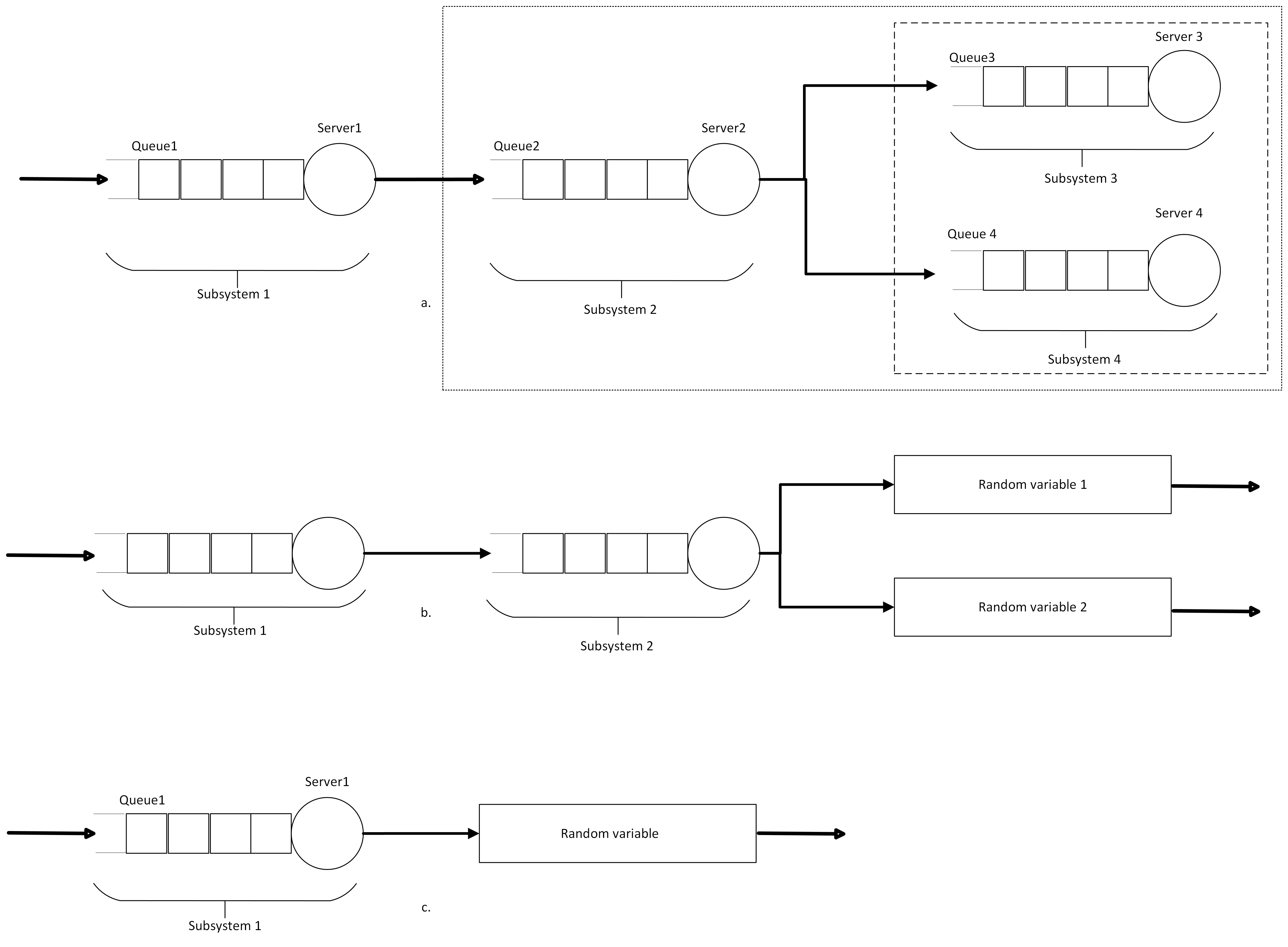}
    \caption{RS prediction validation: the parent model and two corresponding metasimulation models}
    \label{fig:3-stage-val}
\end{figure}

\begin{figure}
    \centering
    \includegraphics[width=0.75\linewidth]{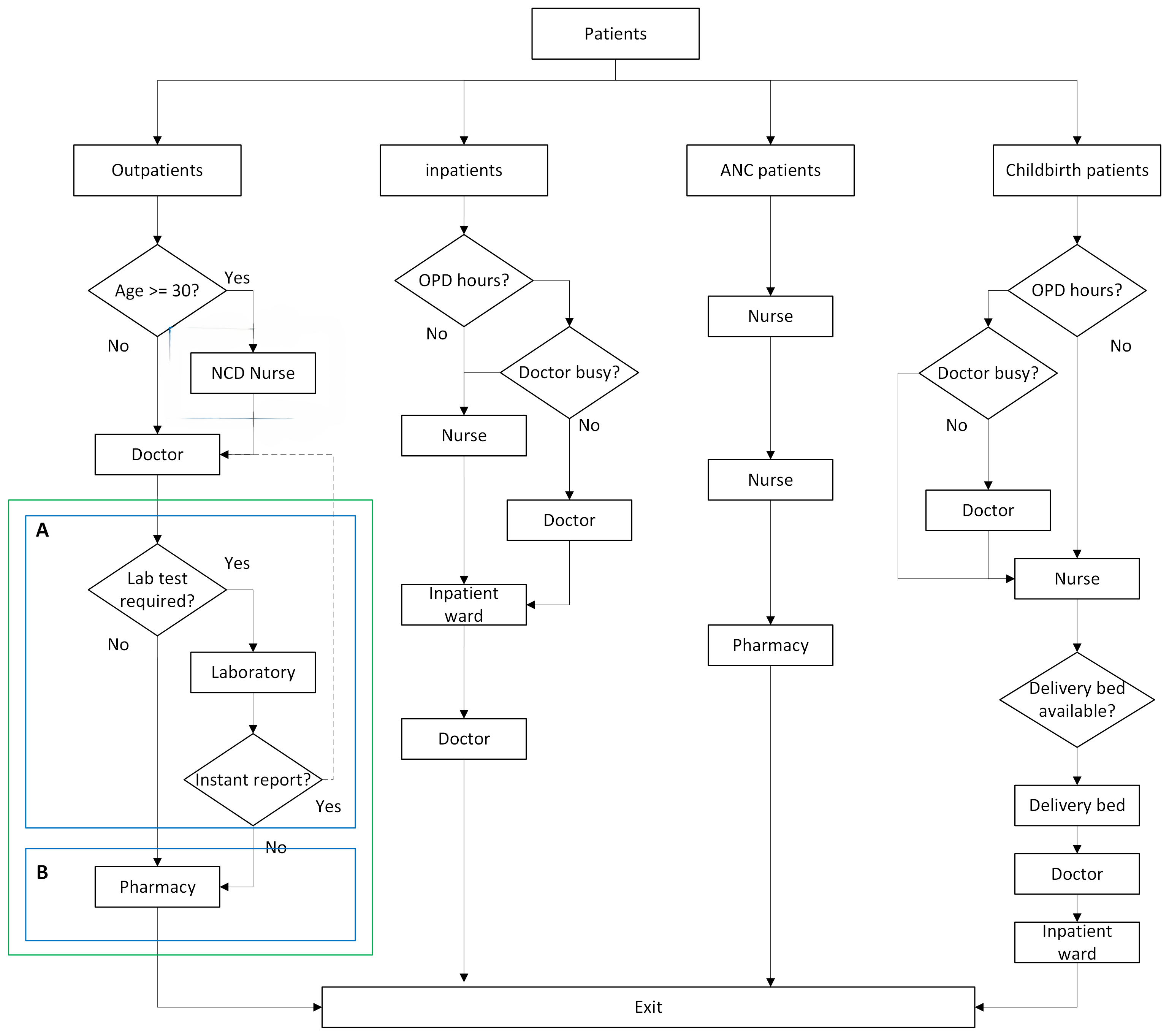}
    \caption{Conceptual model of patient flow in the primary health centre. Source:~\cite{shoaibPHC}. \\
    \emph{Note: Coloured boxes (A,\& B) are used to indicate the subsystems that were abstracted out for validation experiments.}}
    \label{fig:phc-valid}
\end{figure}

We carried out three sets of experiments to demonstrate the deployment of and validate the proposed RS prediction framework. These are listed in Table \ref{tab:valid-exper-summ}. 
The first set of experiments involved applying the RS prediction models for an abstraction operation on a simple two-stage queueing system, as shown in Figure~\ref{figure:abstraction}, where the second stage was abstracted out. For the second set of validation experiments, a more complex three-stage queueing system with two parallel routings in the third stage was conceptualised. This configuration is shown in Figure~\ref{fig:3-stage-val}. For the third set of experiments, a simulation model of the flow of patients in a primary health centre (PHC) developed (displayed in Figure~\ref{fig:phc-valid}) by~\cite{shoaibPHC} was adopted. The code for this PHC simulation was accessed from its \href{https://github.com/shoaibiocl/PHC-}{GitHub} page.

\begin{table}[!htbp]
\centering
\caption{Overview of validation experiments for the runtime savings prediction models.} \label{tab:valid-exper-summ}
\small
\begin{tabular}{|C{2cm}|C{4cm}|C{7.7 cm}|}
\hline
Validation experiment & DES model                                                                             & Description                                                                                    \\ \hline
1                        & 2-stage tandem   queueing system (Figure~\ref{figure:abstraction})                                                      & Scenario: Second   stage abstracted out and replaced with its LOS random variable; $S1$ operation; $O = (\{2\}, \{\}, \{\})$                    \\ \hline
\multirow{2}{*}{2}       & 3-stage tandem queuing system with 2 parallel & Scenario 1:  Stage 3 abstracted out and each subsystem  replaced with its LOS random variable; $S1$ operation, with $O = (\{3,4\}, \{\}, \{\})$  \\ \cline{3-3} 
                            & 
                            subsystems in the third stage (Figure~\ref{fig:3-stage-val}) & Scenario 2:  Stages 2 \& 3   abstracted out by a single LOS random variable; $S_2$ operation, with $O = (\{\}, \{2,3,4\}, \{\})$         \\ \hline
\multirow{3}{*}{3}       &                             & Scenario 1: Laboratory  abstracted out by its random variable; $S1$ operation; $S1$ operation, with $O = (\{Lab\}, \{\}, \{\})$                      \\ \cline{3-3} 
                         &                                                                          Patient flow in a primary health centre (Figure~\ref{fig:phc-valid})      & Scenario 2: Pharmacy abstracted out by its LOS random variable; $S1$ operation, with $O = (\{Pharmacy\}, \{\}, \{\})$                          \\ \cline{3-3} 
                         &                                                                                       & Scenario 3: Laboratory \& pharmacy abstracted out by a single LOS random variable; $S2$ operation, with $O = (\{\}, \{Lab, Pharmacy\}, \{\})$      \\ \hline
\end{tabular}
\end{table}

The approach adopted for the validation experiments is described in detail in Table \ref{tab:val-approach}. Note that while Algorithm~\ref{algo:absrspred} describes the process for deploying the RS predictive framework for a proposed model simplification endeavour, Table \ref{tab:val-approach} describes the systematic procedure to generate both RS predictions as well as ground truth RS estimates for the experiments in Table~\ref{tab:valid-exper-summ} given that the purpose of these experiments was validation in addition to demonstration of the deployment of the RS predictive framework. We provide an overview 
of this procedure below.

Let us consider an example of any DES model comprising $M/G/$ systems (parent model) and its simplified version constructed by replacing a single subsystem within the parent model with its LOS random variable.
First, the parent simulation model is executed for a given run length and a prespecified number of replications are performed. Next, key outcomes such as the average simulation runtime ($t$), the average number of arrivals ($n$), and average occupancy ($\rho$) of the server to be abstracted out are recorded.
Then Algorithm~\ref{algo:absrspred} is deployed - that is, using the equation for an $M/G/$ system from Table~\ref{tab:crr-equations}, the average number of instructions per arrival ($\theta^{mg}_{ss}$) is estimated using $\rho$ from the previous step.  
Following this, the average total number of instructions ($I_{ss}$) is estimated for the subsystem in question within the parent model by combining $n$ with $\theta^{mg}_{ss}$. Subsequently, the expected RNI $\bar{I}$ is estimated using the relation $I_{ss} - (2 \times n)$, and the value obtained is plugged into the appropriate RS prediction equation to estimate the expected RS ($\phi^{mg}$). 

To validate the results, the simplified model is executed under identical conditions and the average computational runtime is calculated ($t'$). The average actual RS is then calculated as $t-t'$ and compared with the predicted RS $\phi^{mg}$. The performance of the prediction is assessed with the mean absolute percentage error (MAPE), mean percentage error (MPE), the root mean squared error (RMSE), and the R-squared ($R^2$) metrics. The MAPE is calculated as $\sum\limits_{j = 1}^{N} \frac{|O_j - P_j|}{O_j} \times 100$, where $O_j$ and $P_j$ are the observed and predicted values of the regression dependent variable, and $N$ represents the size of the sample. Similarly, the MPE is calculated as $\sum\limits_{j = 1}^{N} \frac{(O_j - P_j)}{O_j} \times 100$. The RMSE is calculated as $\sqrt{\sum\limits_{j = 1}^{N} (O_j - P_j)^2}$. The coefficient of determination $R^2$ is a well-established metric to assess the quality of regression fit.

\begin{table}[htbp]
    \centering
\caption{Runtime savings prediction model validation approach.}   
\label{tab:val-approach}
\begin{description}
\item For each server occupancy value of the subsystem to be abstracted out, do: \\
\item \textbf{Step 1: Execute parent model and record data.} 
   \begin{subroutine}
     \item Identify the subsystems to be abstracted out and their queueing disciplines (for example, single-server $G/G/1$ system).
   \item Execute the parent model and record average server occupancy value ($\rho$), number of arrivals ($n$), and computational runtime ($t$).
   \end{subroutine}

\item \textbf{Step 2: Generate runtime savings prediction.}  
   \begin{subroutine}
        \item Call Algorithm~\ref{algo:absrspred} to generate RS prediction $\phi_{pred}$.
   \end{subroutine}

\item \textbf{Step 3: Compare the predicted RS with the actual RS.}
  \begin{subroutine}
  \item Execute the simplified model under conditions identical to that in Step 1(b) to obtain the average computational runtime $t'$.
  \item Calculate the difference in actual and predicted runtimes ($\phi_{act} = t-t'$) to obtain the actual average RS.
  \item Compare $\phi_{pred}$ and $\phi_{act}$ and record data for computing the performance metrics.
  \end{subroutine}
\end{description}
\end{table}

\textbf{Validation experiment 1.} For this set of validation experiments, we worked with the system depicted in Figure~\ref{figure:abstraction} and adhered to the process described in Table \ref{tab:val-approach} for the experiment. The experiments were performed for three cases, wherein the second-stage subsystem (the one being abstracted out) was conceptualised as: (a) an $M/M/1$ system, (b) an $M/G/1$ system, and (c) a $G/G/1$ system. 
The validation was performed for 20 randomly generated utilisation values between 20\% and 93\% for the second-stage subsystem and 30 replications were performed at each utilisation value. 

For both the $M/M/1$ and $M/G/1$ cases, the interarrival times of service-seeking entities, which was varied to generate the desired second stage subsystem occupancy level, were governed by the exponential distribution. With regard to the service time in the $M/G/1$ case, the service time for the first subsystem was modelled as an exponential random variable with mean $1$ minute, and the second subsystem's server had a uniformly distributed service time ranging from $3$ to $9$ minutes. In the $G/G/1$ case, the first subsystem's server had an exponentially distributed service time with a mean of 2 minutes. The second subsystem's server had a normally distributed service time with a mean of 2 minutes and a standard deviation of $0.2$ minutes. The interarrival time of service-seeking entities was modelled using a uniform distribution. The outcomes from the experiments showing actual RS and predicted RS values at different server utilisation values are listed in Table~\ref{tab:2stage-val-2}. Performance metric estimates summarising the performance of the RS prediction approach for all three cases are shown in Table \ref{tab:2stage-val}. 


\begin{table}[htbp]
\small
\centering
\caption{Validation experiment 1 results for the $M/M/1$, $M/G/1$ and $G/G/1$ cases. Note: all runtime savings are in seconds; $\rho$ = second-stage server utilisation level; RS = runtime savings.}
\label{tab:2stage-val-2}
\begin{tabular}{|C{1.2cm}|C{1.2cm}|C{1.2cm}|C{1.2cm}|C{1.2cm}|C{1.2cm}|C{1.2cm}|C{1.2cm}|C{1.2cm}|}

\hline

\multicolumn{3}{|c|}{Abstracted subsystem: $M/M/1$} &\multicolumn{3}{|c|}{Abstracted subsystem: $M/G/1$} & \multicolumn{3}{|c|}{Abstracted subsystem: $G/G/1$} \\ \hline
$\rho$ & Actual RS & Predicted RS & $\rho$ & Actual RS & Predicted RS & $\rho$ & Actual RS & Predicted RS \\ \hline
22.03\%  & 6.06 & 5.53  & 27.52\% & 0.59 & 0.63 & 23.27\% & 1.76 & 1.88 \\ \hline
25.65\%  & 6.24 & 6.48  & 28.22\% & 0.60 & 0.65 & 28.08\% & 2.09 & 2.25 \\ \hline
30.50\%  & 7.99 & 7.77  & 31.05\% & 0.64 & 0.71 & 28.63\% & 2.19 & 2.30 \\ \hline
31.94\%  & 8.25 & 8.16  & 36.99\% & 0.76 & 0.83 & 32.90\% & 2.52 & 2.63 \\ \hline
33.09\%  & 8.72 & 8.46  & 39.23\% & 0.86 & 0.88 & 36.88\% & 2.98 & 2.95 \\ \hline
43.43\%  & 11.02 & 11.27  & 43.13\% & 0.96 & 0.97 & 37.10\% & 2.86 & 2.96 \\ \hline
45.14\%  & 11.83 & 11.76  & 45.11\% & 0.99 & 1.01 & 37.77\% & 2.91 & 3.02 \\ \hline
46.81\%  & 11.65 & 12.21  & 57.28\% & 1.24 & 1.29 & 38.37\% & 2.98 & 3.07 \\ \hline
57.48\%  & 14.51 & 15.22  & 62.82\% & 1.38 & 1.41 & 44.76\% & 3.60 & 3.56 \\ \hline
66.73\%  & 17.72 & 17.86  & 64.66\% & 1.45 & 1.46 & 45.55\% & 3.55 & 3.63 \\ \hline
67.51\%  & 17.27 & 18.11  & 66.78\% & 1.52 & 1.51 & 46.24\% & 3.71 & 3.67 \\ \hline
69.13\%  & 16.71 & 18.58  & 71.90\% & 1.64 & 1.63 & 50.10\% & 3.95 & 3.99 \\ \hline
70.30\%  & 18.09 & 18.93  & 72.03\% & 1.58 & 1.63 & 50.55\% & 3.97 & 4.02 \\ \hline
73.96\%  & 19.41 & 20.02  & 73.53\% & 1.68 & 1.67 & 56.34\% & 4.44 & 4.49 \\ \hline
77.10\%  & 21.72 & 20.94  & 74.59\% & 1.69 & 1.69 & 62.12\% & 5.07 & 4.99 \\ \hline
85.74\%  & 23.64 & 23.54  & 75.92\% & 1.72 & 1.72 & 63.42\% & 5.16 & 5.10 \\ \hline
87.83\%  & 24.15 & 24.18  & 81.78\% & 1.83 & 1.87 & 79.19\% & 6.47 & 6.54 \\ \hline
89.81\%  & 24.34 & 24.79  & 82.83\% & 1.92 & 1.89 & 81.08\% & 6.81 & 6.71 \\ \hline
90.23\%  & 24.37 & 24.92  & 90.04\% & 2.05 & 2.07 & 85.61\% & 7.15 & 7.17 \\ \hline
90.34\%  & 24.48 & 24.95  & 90.24\% & 2.09 & 2.07 & 91.59\% & 7.88 & 7.83 \\ \hline

\end{tabular}
\footnotesize
\textit{ }
\end{table}

\begin{table}[htbp]
    \centering
    \caption{Validation experiment 1 results summary: performance metric values.}
    \label{tab:2stage-val}
    \begin{tabular}{|C{2.5cm}|C{2cm}|C{2cm}|C{2cm}|C{2cm}|}
    \hline
        Second stage subsystem  & MAPE &  MPE & RMSE & R-squared \\ \hline
             
        $M/M/1$ & 3.34\% & -1.30\% & 0.63 & 0.99 \\ \hline     
             
        $M/G/1$ & 2.93\% & -2.43\% & 0.03 & 0.99  \\ \hline
       
        $G/G/1$ & 2.45\%  & -1.66\%  & 0.08 & 0.99 \\ \hline
       
    \end{tabular}
\end{table}

From Tables \ref{tab:2stage-val} and \ref{tab:2stage-val-2}, the performance of the RS prediction approach for all three cases is encouraging - for example, MAPE scores are below 4\% and $R^2$ values are 99\%. 

\textbf{Validation experiment 2.} The three-stage queueing system developed for this validation experiment is depicted in Figure~\ref{fig:3-stage-val}. The parent model is shown at the top of the figure. Two simplification scenarios were considered. For the first scenario, the third stage was abstracted as follows: subsystems 3 and 4 (enclosed in a dashed box in Figure~\ref{fig:3-stage-val}) were each replaced with their corresponding LOS random variables - an $S1$ simplification operation. The corresponding simplified model is depicted in the middle system in Figure~\ref{fig:3-stage-val}. In the second scenario, the second and third stages (enclosed with a dotted box) were jointly abstracted out with a single LOS random variable - an $S2$ operation. The simplified model thus constructed can be seen in Figure~\ref{fig:3-stage-val} (the bottom system). The validation experiment followed the process described in Table~\ref{tab:val-approach}. In this section, we provide detailed results only for the $M/G/$ case, summary results for the performance of the RS prediction method for all cases, and provide detailed results for the $M/M/$ and $G/G/$ cases in the Appendices.

The subsystem service and interarrival times for the $M/G/$ case are as follows. Both subsystems 1 and 2 feature identical servers, each with service time following an exponential distribution with mean 2 minutes. The third and fourth servers, while maintaining the same average service time of 6 minutes and a standard deviation of 2 minutes, differ in their respective distributions. The third subsystem's service times were sampled from a Gaussian distribution, and the fourth server's from a Gamma distribution. The interarrival time of jobs to the system was governed by an exponential distribution whose parameter was calculated based on the desired subsystem server occupancy and subsystem service rate values. The outcomes of the experiments are listed in Table \ref{tab:mg1}.
Now, for the validation experiments, 20 random values were generated between 20\% and 93\% representing the server occupancy of the third subsystem. For these values, the parent model was executed and the number of arrivals, actual server occupancy values, and computational runtimes were recorded. These are listed in the first eight columns in Table~\ref{tab:mg1}. Then, to generate RS predictions, Algorithm~\ref{algo:absrspred} was called for the simplification operation $O = (\{3,4\},\{\},\{\})$. The predicted RS was then compared against the actual runtime recorded on executing the simplified model for each server occupancy value. Overall, the observed average MAPE value calculated across the experiments was 4.66\%, indicating encouraging performance of the RS prediction framework. 

\begin{landscape}
\begin{table}[htbp]
\caption{Validation experiment 2 scenario 1 outcomes comparing predicted and actual runtime savings for the M/G model. Note: $\rho$ = server occupancy; $n$ = average number of arrivals; $s$ = subsystem; $T_p$ = parent model average runtime (seconds); $\theta$ = average number of instructions per arrival; $\bar{I}$ = expected reduction in the number of instructions; APE = absolute percentage error; PE = percentage error} \label{tab:mg1}
\centering
\small
\begin{tblr}{
  width = \linewidth,
  colspec = {|Q[42]|Q[42]|Q[42]|Q[42]|Q[56]|Q[56]|Q[56]|Q[44]|Q[50]|Q[50]|Q[60]|Q[60]|Q[62]|Q[82]|Q[60]|Q[60]|Q[66]|Q[66]|},
  hline{1,22} = {-}{0.08em},
}
$\rho_{s1}$ & $\rho_{s2}$ & $\rho_{s3}$ & $\rho_{s4}$ & $n_{s1}$ \& $n_{s2}$ & $n_{s3}$ & $n_{s4}$ & $T_p$ & $\theta_{s3}$ & $\theta_{s4}$ & $\bar{I}_{s3}$ & $\bar{I}_{s4}$ & $\bar{I}_{s3} + \bar{I}_{s4}$ & Predicted RS & Actual RS & Error  & APE    & PE    \\ \hline
0.17                   & 0.17                   & 0.21                   & 0.31                   & 44927          & 17998               & 26928               & 4.91     & 9.16         & 9.29         & 128864        & 196304        & 325168    & 1.11          & 1.076      & $-0.03$ & 2.90\% & $-2.90\%$ \\\hline
0.19                   & 0.19                   & 0.23                   & 0.35                   & 50634          & 20238               & 30395               & 5.67     & 9.19         & 9.34         & 145585        & 223030        & 368615    & 1.25          & 1.317      & 0.070  & 5.29\% & 5.29\%  \\\hline
0.20                   & 0.20                   & 0.24                   & 0.35                   & 51523          & 20617               & 30906               & 5.76     & 9.20         & 9.35         & 148427        & 226990        & 375417    & 1.27          & 1.318      & 0.049  & 3.71\% & 3.71\%  \\\hline
0.23                   & 0.23                   & 0.28                   & 0.41                   & 60453          & 24158               & 36295               & 6.79     & 9.25         & 9.42         & 175164        & 269144        & 444308    & 1.49          & 1.579      & 0.087  & 5.53\% & 5.53\%  \\\hline
0.24                   & 0.24                   & 0.29                   & 0.44                   & 64262          & 25710               & 38553               & 7.25     & 9.27         & 9.44         & 186978        & 286998        & 473976    & 1.59          & 1.640      & 0.051  & 3.13\% & 3.13\%  \\\hline
0.25                   & 0.25                   & 0.29                   & 0.44                   & 64542          & 25800               & 38742               & 7.30     & 9.27         & 9.45         & 187670        & 288495        & 476165    & 1.60          & 1.715      & 0.119  & 6.96\% & 6.96\%  \\\hline
0.25                   & 0.25                   & 0.30                   & 0.45                   & 65312          & 26120               & 39192               & 7.39     & 9.28         & 9.45         & 190114        & 292060        & 482174    & 1.62          & 1.675      & 0.059  & 3.53\% & 3.53\%  \\\hline
0.28                   & 0.28                   & 0.34                   & 0.51                   & 73997          & 29603               & 44394               & 8.22     & 9.33         & 9.52         & 216896        & 333661        & 550556    & 1.84          & 1.782      & $-0.058$ & 3.27\% & $-3.27\%$ \\\hline
0.29                   & 0.29                   & 0.35                   & 0.52                   & 76265          & 30462               & 45803               & 8.61     & 9.34         & 9.53         & 223550        & 344991        & 568541    & 1.90          & 1.927      & 0.027  & 1.39\% & 1.39\%  \\\hline
0.29                   & 0.29                   & 0.35                   & 0.53                   & 76935          & 30804               & 46130               & 8.78     & 9.34         & 9.54         & 226202        & 347624        & 573826    & 1.92          & 2.092      & 0.174  & 8.33\% & 8.33\%  \\\hline
0.30                   & 0.30                   & 0.36                   & 0.53                   & 77898          & 31186               & 46712               & 8.67     & 9.349         & 9.54         & 229169        & 352327        & 581496    & 1.94          & 1.936      & $-0.007$ & 0.37\% & $-0.37\%$ \\\hline
0.33                   & 0.33                   & 0.40                   & 0.60                   & 87163          & 34850               & 52312               & 9.94     & 9.40         & 9.61         & 257780        & 397864        & 655644    & 2.19          & 2.332      & 0.142  & 6.09\% & 6.09\%  \\\hline
0.33                   & 0.33                   & 0.40                   & 0.60                   & 87894          & 35147               & 52746               & 10.04    & 9.40         & 9.61         & 260112        & 401416        & 661529    & 2.21          & 2.299      & 0.089  & 3.89\% & 3.89\%  \\\hline
0.42                   & 0.42                   & 0.51                   & 0.76                   & 111468         & 44565               & 66903               & 12.88    & 9.52         & 9.75         & 335030        & 518586        & 853616    & 2.86          & 3.043      & 0.181  & 5.94\% & 5.94\%  \\\hline
0.42                   & 0.42                   & 0.51                   & 0.76                   & 111577         & 44657               & 66920               & 12.90    & 9.52         & 9.75         & 335767        & 518727        & 854494    & 2.87          & 2.991      & 0.126  & 4.20\% & 4.20\%  \\\hline
0.43                   & 0.43                   & 0.51                   & 0.77                   & 112550         & 45024               & 67526               & 13.05    & 9.52         & 9.76         & 338708        & 523804        & 862512    & 2.89          & 3.051      & 0.158  & 5.18\% & 5.18\%  \\\hline
0.46                   & 0.46                   & 0.55                   & 0.83                   & 120413         & 48116               & 72297               & 13.94    & 9.56         & 9.8         & 363706        & 563817        & 927523    & 3.12          & 3.327      & 0.208  & 6.27\% & 6.27\%  \\\hline
0.47                   & 0.47                   & 0.57                   & 0.85                   & 123980         & 49535               & 74445               & 14.43    & 9.58         & 9.82         & 375224        & 581893        & 957117    & 3.22          & 3.495      & 0.274  & 7.83\% & 7.83\%  \\ \hline 
\end{tblr}
\end{table}
\end{landscape}

In the second scenario for validation experiment 2, the second and third stages of the parent model are abstracted out and replaced with a single random variable (an $S_2$ operation). The simplified model thus constructed is depicted in Figure~\ref{fig:3-stage-val} (third system from the top). 
With respect to calculation of the parent model data (including runtime), the data for the parent model generated for the first scenario is applicable for this scenario as well. Algorithm~\ref{algo:absrspred} is applied to generate the predicted RS for this simplification operation $O = (\{\},\{3,4\},\{\})$.


Next, the simplified model was executed and the runtimes were recorded. Results from comparing the actual RS values against the predicted RS values are documented in Table ~\ref{tab:app-mgi-child2a}. The results of the experiments are once again encouraging, with a maximum MAPE score of approximately 9.5\% and a minimum $R^2$ value of 91\%.

The calculations and results for the $M/M/$ and the $G/G/$ cases are provided in \ref{app:mm1-valid-outcomes} and \ref{app:gg1-valid-outcomes}.

\begin{table}[htbp]
  \centering
  \caption{Validation experiment 2 outcomes for the $M/M/, M/G/$, and $G/G/$ cases.}
    \begin{tabular}{|C{3.25cm}|C{1.5cm}|C{1.25cm}|C{1.25cm}|C{1.25cm}|C{1.75cm}|}
    \hline
    Queueing disciplines of the simplified systems & Scenario  & MAPE & MPE & RMSE & R-squared\\
    \hline
    \multirow{2}{*}{} $M/M/1$ & 1 & 9.48\%  & -9.48\% & 0.63 & 0.99  \bigstrut\\
\cline{2-6}          & 2 & 3.90\% &  -3.90\% & 0.01 & 0.99 \bigstrut\\
    \hline
    \multirow{2}{*}{} $M/G/1$ & 1 & 4.66\% & 3.93\% & 0.13  & 0.99  \bigstrut\\
\cline{2-6}          & 2 & 5.81 \% & -3.94\% & 0.33  & 0.96 \bigstrut\\
    \hline
    \multirow{2}{*}{} $G/G/1$ &  1 & 7.72\% & 0.19\%  &  0.18 & 0.95 \bigstrut\\
\cline{2-6}          & 2 & 4.55\% & -3.01\%  & 0.31   & 0.91     \bigstrut\\
    \hline
    \end{tabular}%
  \label{tab:3stage-val-summary}%
\end{table}%

\textbf{Validation Experiment 3.} 
We considered three scenarios as part of validation experiment 3, focusing on a DES model that represents patient flow in an Indian primary health centre (PHC). This PHC DES was developed by \cite{shoaibPHC}. PHCs cater to the healthcare needs of multiple types of patients, including (a) those visiting on an outpatient basis, (b) those requiring limited inpatient care, (c) pregnant women visiting for childbirth, and (d) pregnant women visiting for outpatient antenatal care. The majority of patients visit for outpatient care given that the focus of the PHC is on provision of primary care. For example, an average of 60-120 patients are served per day on an outpatient basis, whereas only 1-2 patients per day are served on an inpatient basis.
Given the substantially higher volume of `outpatients', our validation experiments focused on abstracting out subsystems in the outpatient section. The subsystems earmarked for abstraction are enclosed within coloured boxes and marked with letters A and B as shown in Figure~\ref{fig:phc-valid}. The first scenario considered the abstraction of the laboratory subsystem that consisted of a queue and a server (a laboratory technician / phlebotomist), marked by a blue box labelled `A' in Figure~\ref{fig:phc-valid}. The second experiment focused on abstracting the pharmacy alone, indicated by a blue box marked as `B'. The pharmacy subsystem consisted of a queue and a pharmacist (server). The third experiment combined the abstraction of both the laboratory and pharmacy subsystems. 

The validation experiments here were conducted for three sets of input parameters. We varied the interarrival rates and maintained the original settings for all the other parameters - such as service rates, number of servers at each subsystem etc. - described in \cite{shoaibPHC}. However, we performed the following modifications to make the DES more amenable for model simplification.
\begin{itemize}
    \item The time allocated to administrative work (e.g., paperwork for patients) for the NCD nurse, staff (inpatient/childbirth) nurse, and doctor was removed from the model. This adjustment was necessary because administrative work time was sampled from a Gaussian distribution in the original PHC model and included in the total activity time of these resources. This would artificially inflate server occupancy levels without influencing the events in the event list or the number of instructions processed by the simulation engine.
    \item The simulation run length was set to 10 years of simulation time.
    
    \item The original model generated outcomes in two separate MS Excel spreadsheets once the model completed its run. We consolidated the results into a single file.
     
\end{itemize}

We replicated the process in Table \ref{tab:val-approach} for the validation experiment. The parent model was executed for 10 replications. The average computational runtime per replication recorded varied between 150 and 235 seconds for the selected interarrival rate values as listed in Table~\ref{tab:phc-valid-outcome}. Following this, per Algorithm~\ref{algo:absrspred}, the RIE was estimated for each case and then expected RS predictions were made using the equation for $\phi_{gg}$. Then the simplified models for each scenario were executed and the actual RS achieved was computed. The outcomes for each scenario are presented in Table~\ref{tab:phc-valid-outcome}. The results reveal a consistent pattern of underestimation of the actual RS by the RS prediction model, which implies that there may be factors yielding more RS than those accounted for in our framework, especially for DES models such as the PHC model. We leave the investigation and incorporation of these factors into our RS prediction framework for future work. That said, we note that the maximum PE is 18.45\%, with an average PE of approximately 15\%, indicating that the prediction performance remains acceptable. Further, the $R^2$ value remains at 99\%, indicating clearly that the RS prediction framework has clearly identified the pattern in the generation of RS from model simplification operations. 

Overall, from the results of validation experiments 1 through 3, our RS prediction approach appears to yield acceptable results.

\begin{table}[!htbp]
\centering
\caption{Validation experiment 3 outcomes for the primary health centre DES model. \textit{Notes. All runtimes shown are mean runtimes (in seconds), with standard deviations in parentheses. RS = runtime savings; PE = percentage error,  $\rho_l$ = laboratory utilisation level, $\rho_p$ = pharmacy utilisation level.}}
\label{tab:phc-valid-outcome}
\small
\setlength{\extrarowheight}{3pt}
\begin{tabular}{|c|l|l|l|l|}
\hline
Experiment  & Outcomes  & Laboratory (A) & Pharmacy (B) & Laboratory \& \\
            &            &               &               & Pharmacy (A+B) 
\\ \hline
\multirow{6}{*}{\begin{tabular}[c]{@{}c@{}}Set 1\\           $ \rho_{l} $ = 46.60\%\\      $ \rho_{p} $ = 53.19\%\end{tabular}}                       
 & Parent model runtime & \multicolumn{3}{c|}{187.18 (1.53)}                                                                \\ \cline{2-5} 
 & Simplified model runtime  & 168.81 (1.09)    & 150.23 (1.15)   & 125.97  (0.63)      \\ \cline{2-5}   
 & Actual RS      & 18.37            & 36.95           & 61.21                \\ \cline{2-5}                           
 & Predicted RS   & 15.23            & 30.6            & 49.91                \\ \cline{2-5}                  
 & Error          & 3.14             & 6.35            & 11.29                 \\ \cline{2-5} 
 & PE             & 17.12\%          & 17.18\%         & 18.45\%                \\ \hline
 
\multirow{6}{*}{\begin{tabular}[c]{@{}c@{}}Set 2\\         $ \rho_{l} $ = 57.57\%\\      $ \rho_{p} $ = 66.42\%\end{tabular}}                     
 & Parent model runtime  & \multicolumn{3}{c|}{235.56 (3.09)}                                                                \\ \cline{2-5} 
 & Simplified model runtime  & 212.62 (1.34)   & 192.72 (1.82)        & 163.51 (1.81)  \\ \cline{2-5} 
 & Actual RS      & 22.94           & 42.83    & 72.05                    \\ \cline{2-5}  
 & Predicted RS     & 19.26  & 39.06    & 63.73    \\ \cline{2-5} 
 & Error          & 3.68            & 3.77     & 8.32                    \\ \cline{2-5} 
 & PE                  & 16.04\%    & 8.80\%  & 11.55\%    \\ \hline

\multirow{6}{*}{\begin{tabular}[c]{@{}c@{}}Set 3\\       $ \rho_{l} $ = 37.70\%\\      $ \rho_{p} $ = 42.52\%\end{tabular}} 
& Parent model runtime  & \multicolumn{3}{c|}{152.11 (1.74)} \\ \cline{2-5} 
& Simplified model runtime & 137.56 (2.06)     & 123.25 (1.45)   & 105.16 (1.00)                   \\ \cline{2-5} 
& Actual RS       & 14.55      & 28.85    & 46.95                    \\ \cline{2-5} 
& Predicted RS     & 12.18      & 24.28    & 39.92                    \\ \cline{2-5} 
& Error            & 2.37       & 4.57     & 7.02                     \\ \cline{2-5} 
& PE                  & 16.28\%    & 15.85\%  & 14.96\%                  \\ \hline
\multicolumn{5}{|c|}{MPE = 15.14\%; RMSE = 6.24; R-squared = 0.99} \\ \hline

\end{tabular}

\end{table}

\section{Discussion and Conclusion}
\label{sec:conc}

This study undertakes a methodological investigation into the simplification of DES models, introducing a queuing-theoretic approach designed to predict computational runtime savings from model simplification through the widely used abstraction and aggregation operations. The paper develops predictive models that estimate potential computational savings from model simplification prior to its actual development, and potentially without needing to run the parent model even once. This is the first study to address this, as previous work on simulation model simplification has primarily focused on model simplification methods themselves via different approaches \citep{Lidberg2021, van2017approaches}.
Thus, we address a gap in existing literature, offering a tool for modellers to pre-emptively assess the benefit from simplification efforts.

In this paper, we discuss in detail the methodology behind the development of the RS prediction models. Our research highlights the key determinants of computational runtime for DESs and use these to develop the predictive framework for estimating RS via model simplification. The RS prediction framework is then validated through a series of computational experiments involving multiple systems of varying complexity, including a set of experiments involving a DES of a real-world healthcare facility. 
The results from the validation experiments are encouraging, thereby indicating the effectiveness of our approach and providing a way forward towards its eventual adoption into both commercial and open-source DES packages.

A limitation of this study is that it was designed and conducted using a particular programming language and DES package. Therefore, the specific equations derived as part of the RS prediction framework may not find direct applicability to other modelling paradigms. However, the study developed and demonstrated a general roadmap that can be replicated with other DES packages or programming languages. We also note that we have not considered DES models of queueing systems with more complex behaviour such as reneging or balking behaviour, or systems with batch arrivals, or stochastic arrival and service patterns, etc. While the fact that our predictive framework relies primarily on server occupancy implies that the approach may be extended with minimal effort to handle behaviours such as balking / reneging, a natural avenue of further research involves formally investigating extension of our approach to such systems.

We envision that this runtime (savings) prediction framework could be integrated into commercial or open-source DES platforms as follows. The generation of the specific equations within the RS predictive framework can be performed as part of the installation process of the simulation platform, given that the model development process does not rely on a specific DES, but only DESs of \textit{2s, ms} and \textit{ss} queueing systems. Consider the scenario of a modeller engaged in conceptualising a model simplification operation for a parent DES comprising a network of queueing subsystems, with attendant subsystem occupancy estimates. The modeller could select the subsystems meant for a specific simplification operation and click on a button that estimates the RS from the simplification operation using our predictive framework. We note here that the RS estimates generated by our proposed framework are in absolute terms - i.e., in units of time; however, the models can be modified with minimal effort to generate relative runtime savings estimates. 


\bibliographystyle{elsarticle-num-names} 
\bibliography{cas-refs}


\appendix

\section{Example Outputs from Parent and Simplified Models}
\label{app:asbtraction}

\begin{figure}[H]
    \centering 
    \begin{subfigure}{\linewidth}
        \centering
        \includegraphics[width=\linewidth]{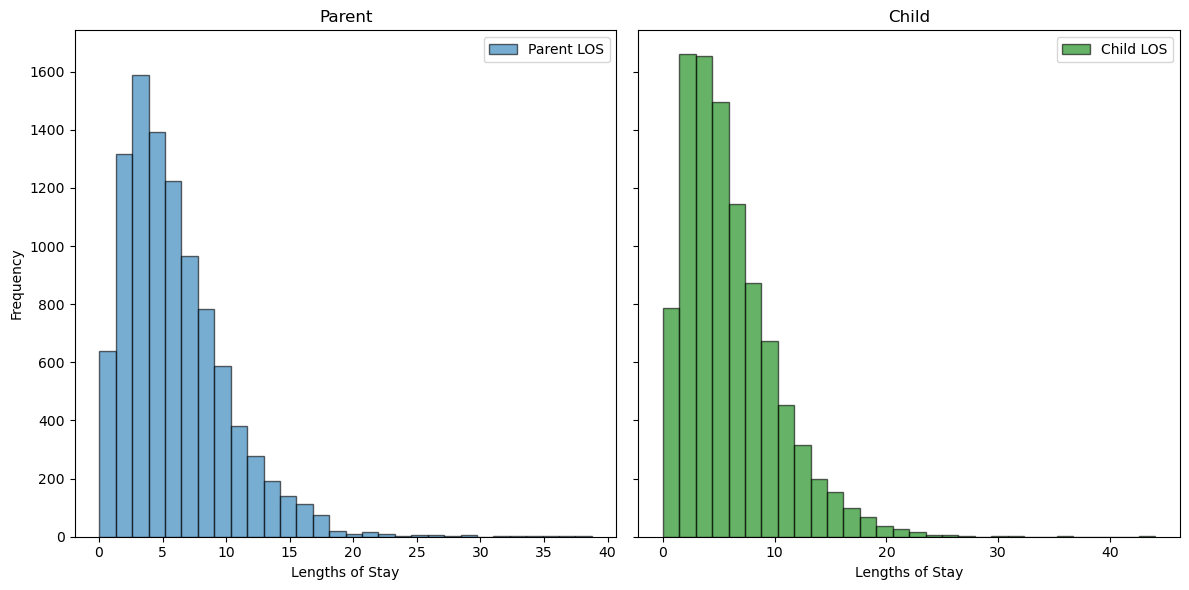} 
        \caption{$M/M/$}
        \label{mm-hist} 
    \end{subfigure}

    \bigskip 

    \begin{subfigure}{\linewidth}
        \centering
        \includegraphics[width=\linewidth]{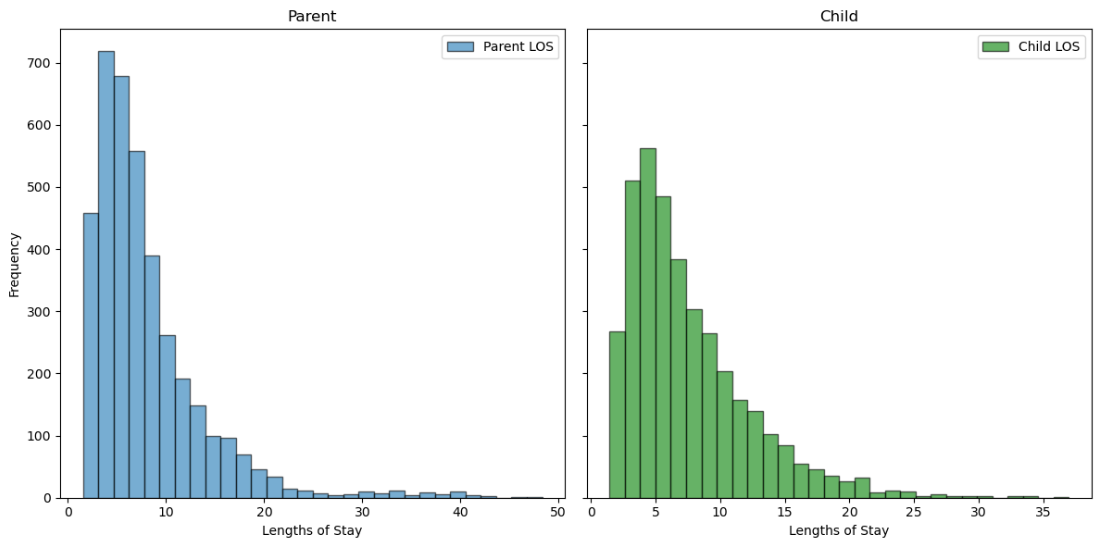} 
        \caption{$M/G/$}
        \label{mg-hist} 
    \end{subfigure}

     \bigskip

    \begin{subfigure}{\linewidth}
        \centering
        \includegraphics[width=\linewidth]{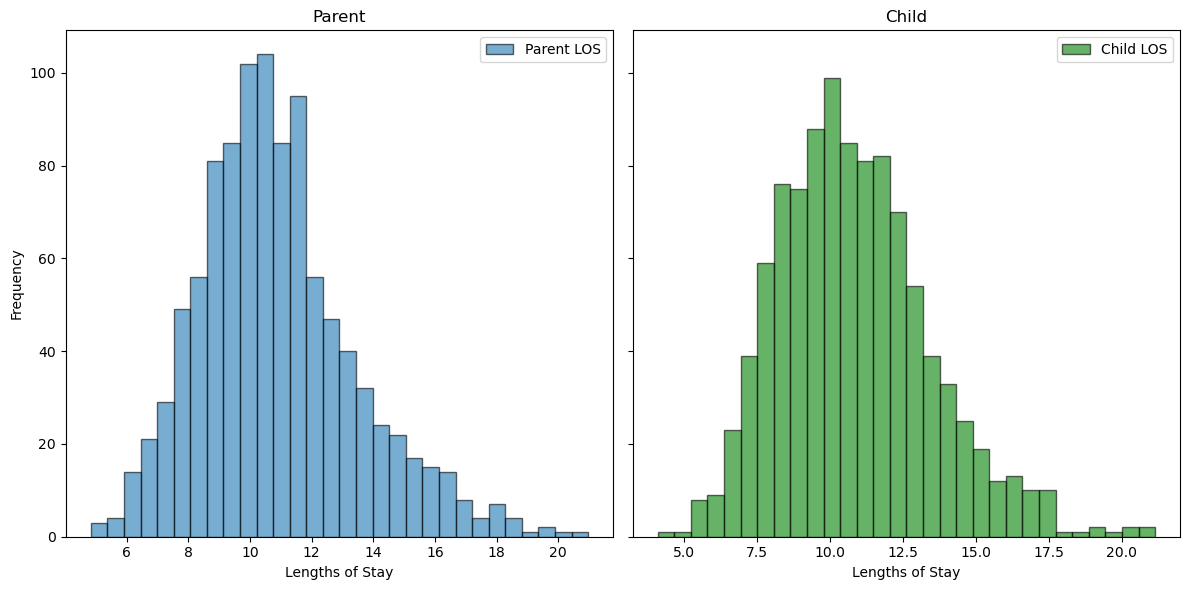} 
        \caption{$G/G/$}
        \label{gg-hist} 
    \end{subfigure}

    \caption{Distributions of lengths of stay data from parent and simplified (child) models.}
    \label{fig:hist} 
\end{figure}

\newpage

\section{Validation Experiment 2: Additional Results}
\label{app:3-stage-val}

\subsection{Additional validation experiment 2 results for the $M/G/$ case.}
\label{app:mg1-valid-outcomes}

\begin{table}[htbp]
  \centering
  \caption{Validation experiment 2, Scenario 2 results: the average number of instructions estimates for the abstracted subsystems 2, 3, \& 4 for the simplified $M/G/$ case.}
  \renewcommand{\arraystretch}{1.1}
  \label{tab:app-mg1-child2}
  \footnotesize
    \begin{tabular}{|p{1 cm}|p{1 cm}|p{1 cm}|p{2.5cm}|p{2cm}|p{2cm}|}
    \hline
    $\theta_{ss2}$ & $\theta_{ss3}$ & $\theta_{ss4}$ & RIE S2 ($I_2 = n_2\times(\theta_{ss2}-2))$ & RIE S3 ($I_3 = n_3 \times \theta_{ss3}$) & RIE S4 ($n_4\times \theta_{ss4}$)  \bigstrut\\
    \hline
    9.26  & 9.27  & 9.45  & 468656 & 239269 & 365980  \bigstrut\\
    \hline
    9.26  & 9.28  & 9.45  & 474413 & 242353 & 370443  \bigstrut\\
    \hline
    9.30  & 9.33  & 9.52  & 539877 & 276101 & 422449 \bigstrut\\
    \hline
    9.30  & 9.34  & 9.53  & 557051 & 284474 & 436597  \bigstrut\\
    \hline
    9.31  & 9.34  & 9.54  & 562157 & 287809 & 439884  \bigstrut\\
    \hline
    9.31  & 9.35  & 9.54  & 569481 & 291541 & 445750  \bigstrut\\
    \hline
    9.35  & 9.40  & 9.61  & 640225 & 327480 & 502488  \bigstrut\\
    \hline
    9.35  & 9.40  & 9.61  & 645814 & 330406 & 506909  \bigstrut\\
    \hline
    9.44  & 9.52  & 9.75  & 828809 & 424161 & 652392  \bigstrut\\
    \hline
    9.44  & 9.52  & 9.75  & 829662 & 425081 & 652567  \bigstrut\\
    \hline
    9.44  & 9.52  & 9.76  & 837383 & 428756 & 658856  \bigstrut\\
    \hline
    9.47  & 9.56  & 9.80  & 899349 & 459937 & 708412  \bigstrut\\
    \hline
    7.48  & 9.57  & 9.82  & 679717 & 474294 & 730782  \bigstrut\\
    \hline
    
    \end{tabular}%
    
    \textit{Note: RIE S2 stands for the reduction in the number of instructions due to subsystem 2.}
  
\end{table}%

\begin{table}[htbp]
  \centering
  \caption{Validation experiment 2, Scenario 2 results: comparing expected and actual runtime savings for the simplified $M/G/$ case.}
    \label{tab:app-mgi-child2a}%
  \footnotesize
  \renewcommand{\arraystretch}{1.1}
    \begin{tabular}{|C{1.5cm}|C{3cm}|C{1.5cm}|C{3cm}|C{3cm}|}
    \hline
    Predicted Runtime Savings & Actual Runtime Savings & Error & Percentage error (PE)   & Absolute PE \bigstrut\\
    \hline
3.85         & 3.60      & -0.25 & -6.98\%  & 6.98\%  \\ \hline
3.89         & 3.76      & -0.13 & -3.57\%  & 3.57\%  \\ \hline
4.44         & 4.57      & 0.12  & 2.69\%   & 2.69\%  \\ \hline
4.59         & 4.20      & -0.39 & -9.20\%  & 9.20\%  \\ \hline
4.63         & 4.35      & -0.28 & -6.51\%  & 6.51\%  \\ \hline
4.69         & 4.21      & -0.48 & -11.50\% & 11.50\% \\ \hline
5.29         & 5.35      & 0.06  & 1.05\%   & 1.05\%  \\ \hline
5.34         & 5.02      & -0.32 & -6.36\%  & 6.36\%  \\ \hline
6.91         & 6.42      & -0.48 & -7.50\%  & 7.50\%  \\ \hline
6.91         & 6.46      & -0.45 & -6.96\%  & 6.96\%  \\ \hline
6.98         & 6.66      & -0.32 & -4.80\%  & 4.80\%  \\ \hline
7.52         & 7.68      & 0.16  & 2.08\%   & 2.08\%  \\ \hline
6.77         & 7.22      & 0.45  & 6.30\%   & 6.30\%  \\ \hline
    
    \end{tabular}%
  \label{tab:app-mgi-child2a}%
  
\end{table}%


\subsection{Validation experiment 2: results for the $M/M/$ case.}
\label{app:mm1-valid-outcomes}

\begin{table}[htbp]
  \centering
 \caption{Validation experiment 2, Scenario 1 results: parent simulation model outcomes - server occupancy values, number of arrivals, and runtime for the $M/M/$ case.  }
\label{tab:app-mm1-parent}
   \footnotesize
\begin{tabular}{|p {1.5 cm}|p{1.5 cm}|p{1.5 cm}|p{1.5 cm}|p{1.65 cm}|p{1.8 cm}|p{1.8 cm}|p{1cm}|}
    \hline
    Server 1 utilisation & Server 2 utilisation & Server 3 utilisation & Server 4 utilisation & Arrivals in subsystem 1 \& 2 ($n_2$) & Arrivals in subsystem 3 ($n_3$) & Arrivals subsystem 4 ($n_4$) & Run time (s) \\
    \hline
    0.13  & 0.19  & 0.18  & 0.19  & 34159 & 13654 & 20504 & 3.76 \bigstrut\\
    \hline
    0.16  & 0.24  & 0.22  & 0.24  & 42038 & 16835 & 25204 & 4.71 \bigstrut\\
    \hline
    0.20  & 0.30  & 0.28  & 0.30  & 52467 & 20978 & 31488 & 5.83 \bigstrut\\
    \hline
    0.23  & 0.34  & 0.32  & 0.34  & 60374 & 24154 & 36221 & 6.73 \bigstrut\\
    \hline
    0.25  & 0.37  & 0.35  & 0.38  & 65722 & 26289 & 39433 & 7.37 \bigstrut\\
    \hline
    0.26  & 0.39  & 0.36  & 0.39  & 68237 & 27300 & 40937 & 7.67 \bigstrut\\
    \hline
    0.29  & 0.43  & 0.41  & 0.43  & 76090 & 30451 & 45639 & 8.60 \bigstrut\\
    \hline
    0.30  & 0.45  & 0.42  & 0.45  & 78708 & 31487 & 47222 & 8.91 \bigstrut\\
    \hline
    0.31  & 0.46  & 0.43  & 0.46  & 81271 & 32505 & 48766 & 9.21 \bigstrut\\
    \hline
    0.32  & 0.48  & 0.45  & 0.48  & 83938 & 33511 & 50426 & 9.53 \bigstrut\\
    \hline
    0.33  & 0.49  & 0.46  & 0.49  & 86448 & 34580 & 51868 & 9.84 \bigstrut\\
    \hline
    0.33  & 0.50  & 0.47  & 0.50  & 87833 & 35129 & 52703 & 9.99 \bigstrut\\
    \hline
    0.33  & 0.50  & 0.46  & 0.50  & 87206 & 34914 & 52293 & 9.92 \bigstrut\\
    \hline
    0.37  & 0.55  & 0.51  & 0.55  & 96039 & 38374 & 57665 & 10.97 \bigstrut\\
    \hline
    0.40  & 0.60  & 0.56  & 0.60  & 104983 & 42010 & 62972 & 11.99 \bigstrut\\
    \hline
    0.43  & 0.64  & 0.60  & 0.64  & 111715 & 44678 & 67036 & 12.95 \bigstrut\\
    \hline
    0.44  & 0.66  & 0.62  & 0.66  & 115472 & 46206 & 69266 & 13.39 \bigstrut\\
    \hline
    0.46  & 0.69  & 0.64  & 0.69  & 120289 & 48174 & 72116 & 13.99 \bigstrut\\
    \hline
    0.47  & 0.71  & 0.66  & 0.71  & 124019 & 49600 & 74418 & 14.45 \bigstrut\\
    \hline
    \end{tabular}%
  \label{tab:addlabel}%
\end{table}%

\begin{table}[htbp]
  \caption{Validation experiment 2, Scenario 1 results: the number of instructions estimated for the abstracted subsystems 3 and 4, with comparisons between expected and actual runtime savings for the simplified $M/M/$ case.}
  \footnotesize
  \renewcommand{\arraystretch}{1.1}
    \begin{tabular}{|p {0.6 cm}|p{0.6 cm}|p{1.5 cm}|p{1.5 cm}|p{1.25 cm}|p{1.35 cm}|p{1
 cm}|p{1 cm}|p{1.2 cm}|p{1.2 cm}|}
    \hline
   $\theta_{ss3}$ & $\theta_{ss4}$& RNI S3 ($I_3 = n_3\times(\theta_{ss3}-2))$ & RNI S4  ($I_4 =n_4\times(\theta_{ss4}-2))$  & Total RNI $(I_3 + I_4)$  & Predicted RS & Actual RS & Error & PE   & APE \\ \hline
    9.30 & 9.31 & 153023 & 230286 & 383309 & 1.44  & 1.34  & -0.11 & -7.94\% & 7.94\% \\ \hline
    9.34 & 9.36 & 177166 & 266492 & 443657 & 1.68  & 1.56  & -0.12 & -7.87\% & 7.87\% \\ \hline
    9.36 & 9.39 & 193574 & 291328 & 484902 & 1.84  & 1.69  & -0.15 & -9.11\% & 9.11\% \\ \hline
    9.38 & 9.40 & 201393 & 302979 & 504372 & 1.92  & 1.74  & -0.17 & -9.95\% & 9.95\% \\ \hline
    9.42 & 9.45 & 225897 & 339800 & 565697 & 2.16  & 1.95  & -0.20 & -10.40\% & 10.40\% \\ \hline
    9.43 & 9.46 & 233992 & 352290 & 586282 & 2.24  & 1.98  & -0.26 & -13.16\% & 13.16\% \\ \hline 
    9.44 & 9.47 & 241978 & 364497 & 606475 & 2.32  & 2.11  & -0.20 & -9.69\% & 9.69\% \\ \hline
    9.46 & 9.49 & 249932 & 377682 & 627613 & 2.40  & 2.18  & -0.22 & -9.99\% & 9.99\% \\ \hline
    9.47 & 9.50 & 258370 & 389153 & 647523 & 2.48  & 2.27  & -0.20 & -9.00\% & 9.00\% \\ \hline
    9.48 & 9.51 & 262702 & 395840 & 658542 & 2.52  & 2.31  & -0.21 & -9.10\% & 9.10\% \\ \hline
    9.48 & 9.51 & 260991 & 392598 & 653590 & 2.50  & 2.27  & -0.23 & -10.21\% & 10.21\% \\ \hline
    9.52 & 9.56 & 288626 & 435834 & 724460 & 2.78  & 2.56  & -0.22 & -8.63\% & 8.63\% \\ \hline
    9.57 & 9.61 & 317926 & 479028 & 796953 & 3.06  & 2.69  & -0.37 & -13.74\% & 13.74\% \\ \hline
    9.60 & 9.65 & 339713 & 512475 & 852189 & 3.27  & 3.02  & -0.25 & -8.36\% & 8.36\% \\ \hline
    9.62 & 9.67 & 352228 & 530929 & 883157 & 3.39  & 3.21  & -0.19 & -5.84\% & 5.84\% \\ \hline
    9.65 & 9.70 & 368462 & 554774 & 923236 & 3.55  & 3.25  & -0.30 & -9.11\% & 9.11\% \\ \hline
    9.67 & 9.71 & 380328 & 574038 & 954366 & 3.67  & 3.37  & -0.30 & -9.05\% & 9.05\% \\ \hline
    \end{tabular}%
  \label{tab:mm-child1-valid}%
\end{table}%

\begin{table}[htbp]
  \centering
  \centering
  \caption{Validation experiment 2, Scenario 2 results: number of instructions estimated for the abstracted subsystems 2, 3, \& 4 for the child $M/M/$ case.}
  \label{tab:app-mm1-child2}
  \footnotesize
    \begin{tabular}{|p{1 cm}|p{1 cm}|p{1 cm}|p{2.5cm}|p{2cm}|p{2cm}|}
    \hline
        $\theta_{ss2}$ & $\theta_{ss3}$ & $\theta_{ss4}$ & RNI S2 ($I_2 = n_2\times(\theta_{ss2}-2))$ & RNI S3 ($I_3 = n_3 \times \theta_{ss3}$) & RNI S4 ($I_4 = n_4\times \theta_{ss4}$)  \bigstrut\\
    \hline
    9.36  & 9.33  & 9.36  & 444105 & 177166 & 266492 \bigstrut\\
    \hline
    9.39  & 9.36  & 9.39  & 485241 & 193574 & 291328 \bigstrut\\
    \hline
    9.40  & 9.38  & 9.40  & 504802 & 201393 & 302979 \bigstrut\\
    \hline
    9.45  & 9.42  & 9.45  & 567190 & 225897 & 339800 \bigstrut\\
    \hline
    9.46  & 9.43  & 9.46  & 587050 & 233992 & 352290 \bigstrut\\
    \hline
    9.47  & 9.44  & 9.47  & 607248 & 241978 & 364497 \bigstrut\\
    \hline
    9.49  & 9.46  & 9.49  & 629310 & 249932 & 377682 \bigstrut\\
    \hline
    9.50  & 9.47  & 9.50  & 649402 & 258370 & 389153 \bigstrut\\
    \hline
    9.51  & 9.48  & 9.51  & 660217 & 262702 & 395840 \bigstrut\\
    \hline
    9.51  & 9.48  & 9.51  & 654894 & 260991 & 392598 \bigstrut\\
    \hline
    9.56  & 9.52  & 9.56  & 725207 & 288626 & 435834 \bigstrut\\
    \hline
    9.61  & 9.57  & 9.61  & 798429 & 317926 & 479028 \bigstrut\\
    \hline
    9.64  & 9.60  & 9.64  & 853626 & 339713 & 512475 \bigstrut\\
    \hline
    9.66  & 9.62  & 9.67  & 884191 & 352228 & 530929 \bigstrut\\
    \hline
    9.69  & 9.65  & 9.69  & 926378 & 368462 & 554774 \bigstrut\\
    \hline
    9.71  & 9.67  & 9.71  & 957053 & 380328 & 574038 \bigstrut\\
    \hline
    \end{tabular}%

\end{table}%

\begin{table}[htbp]
  \centering
  
  \caption{Validation experiment 2, Scenario 2 results: comparing expected and actual runtime savings for the simplified $M/M/$ case.}
    \label{tab:app-mm1-child2a}%
  \footnotesize
    \begin{tabular}{|p{2.3 cm}|p{2 cm}|p{1 cm}|p{1.5 cm}|p{1.5 cm}|p{1.5
 cm}|}
    \hline
    Predicted RS & Actual RS & Error & PE & APE \bigstrut\\
    \hline
   
    \hline
    3.41  & 3.26  & -0.15 & -4.66\% & 4.66\% \bigstrut\\
    \hline
    3.73  & 3.60  & -0.13 & -3.63\% & 3.63\% \bigstrut\\
    \hline
    3.89  & 3.73  & -0.15 & -4.04\% & 4.04\% \bigstrut\\
    \hline
    4.37  & 4.19  & -0.18 & -4.34\% & 4.34\% \bigstrut\\
    \hline
    4.53  & 4.35  & -0.18 & -4.13\% & 4.13\% \bigstrut\\
    \hline
    4.68  & 4.51  & -0.17 & -3.85\% & 3.85\% \bigstrut\\
    \hline
    4.85  & 4.67  & -0.18 & -3.94\% & 3.94\% \bigstrut\\
    \hline
    5.01  & 4.81  & -0.20 & -4.17\% & 4.17\% \bigstrut\\
    \hline
    5.09  & 4.89  & -0.21 & -4.23\% & 4.23\% \bigstrut\\
    \hline
    5.05  & 4.86  & -0.20 & -4.02\% & 4.02\% \bigstrut\\
    \hline
    5.60  & 5.41  & -0.20 & -3.65\% & 3.65\% \bigstrut\\
    \hline
    6.17  & 5.87  & -0.30 & -5.07\% & 5.07\% \bigstrut\\
    \hline
    6.60  & 6.39  & -0.21 & -3.36\% & 3.36\% \bigstrut\\
    \hline
    6.84  & 6.64  & -0.20 & -3.01\% & 3.01\% \bigstrut\\
    \hline
    7.16  & 6.96  & -0.20 & -2.86\% & 2.86\% \bigstrut\\
    \hline
    7.40  & 7.16  & -0.24 & -3.38\% & 3.38\% \bigstrut\\
    \hline
    \end{tabular}%
  \label{tab:addlabel}%
\end{table}%


\subsection{Validation experiment 2: results for the $G/G/$ case.}
\label{app:gg1-valid-outcomes}

\begin{table}[!ht]
    \centering
    \footnotesize
  \caption{Validation experiment 2, Scenario 1: parent simulation model outcomes showing server occupancy values, number of arrivals, and runtimes for the $G/G/$ case.  }
    \label{tab:app-gg1-parent}
   \begin{tabular}{|p {1.5 cm}|p{1.5 cm}|p{1.5 cm}|p{1.5 cm}|p{1.65 cm}|p{1.8 cm}|p{1.8 cm}|p{1cm}|}
    \hline
    Server 1 utilisation & Server 2 utilisation & Server 3 utilisation & Server 4 utilisation & Arrivals in subsystem 1 \& 2 ($n_2$) & Arrivals in subsystem 3 ($n_3$) & Arrivals subsystem 4 ($n_4$) & Runtime (s) \\ \hline
        0.32 & 0.25 & 0.22 & 0.39 & 56603 & 22639 & 33964 & 6.43 \\ \hline
        0.35 & 0.28 & 0.24 & 0.43 & 62198 & 24886 & 37312 & 6.89 \\ \hline
        0.37 & 0.29 & 0.25 & 0.45 & 64989 & 26012 & 38976 & 7.03 \\ \hline
        0.39 & 0.31 & 0.26 & 0.47 & 68851 & 27537 & 41314 & 7.86 \\ \hline
        0.40 & 0.31 & 0.27 & 0.47 & 69382 & 27798 & 41583 & 7.93 \\ \hline
        0.43 & 0.33 & 0.29 & 0.52 & 75360 & 30104 & 45256 & 8.65 \\ \hline
        0.45 & 0.35 & 0.30 & 0.53 & 77971 & 31162 & 46810 & 8.94 \\ \hline
        0.45 & 0.35 & 0.30 & 0.53 & 78131 & 31285 & 46846 & 8.97 \\ \hline
        0.45 & 0.35 & 0.30 & 0.54 & 78658 & 31406 & 47252 & 9.02 \\ \hline
        0.48 & 0.37 & 0.32 & 0.58 & 84452 & 33786 & 50666 & 9.31 \\ \hline
        0.49 & 0.38 & 0.33 & 0.59 & 86001 & 34386 & 51615 & 9.53 \\ \hline
        0.49 & 0.38 & 0.33 & 0.59 & 86360 & 34528 & 51832 & 9.64 \\ \hline
        0.50 & 0.39 & 0.33 & 0.60 & 87104 & 34810 & 52294 & 9.98 \\ \hline
        0.53 & 0.41 & 0.36 & 0.64 & 93207 & 37278 & 55929 & 10.84 \\ \hline
        0.53 & 0.42 & 0.36 & 0.64 & 93729 & 37531 & 56198 & 10.87 \\ \hline
        0.62 & 0.49 & 0.42 & 0.75 & 109410 & 43749 & 65660 & 12.89 \\ \hline
        0.63 & 0.49 & 0.42 & 0.75 & 109580 & 43795 & 65785 & 12.86 \\ \hline
        0.63 & 0.49 & 0.42 & 0.75 & 110147 & 44051 & 66095 & 12.93 \\ \hline
        0.66 & 0.51 & 0.44 & 0.79 & 115298 & 46048 & 69250 & 13.54 \\ \hline
        0.67 & 0.52 & 0.45 & 0.81 & 117705 & 47122 & 70584 & 13.85 \\ \hline
    \end{tabular}
\end{table}

\begin{table}[!ht]
  \centering
  \caption{Validation experiment 2, Scenario 1 results: number of instructions estimated for the abstracted subsystems 3 and 4; and  comparisons between expected and actual runtime savings for the simplified $G/G/$ case.}
  \footnotesize
    \begin{tabular}{|p {0.6 cm}|p{0.6 cm}|p{1.5 cm}|p{1.5 cm}|p{1.25 cm}|p{1.25 cm}|p{1.25 cm}|p{1 cm}|p{1.3 cm}|p{1.1 cm}|}
    \hline
   $\theta_{ss3}$ & $\theta_{ss4}$& RNI S3 ($n_3\times(\theta_{ss3}-2)$ & RNI S4  ($n_4\times(\theta_{ss4}-2))$  & Total RNI $(I_3 + I_4)$  & Predicted RS & Actual RS & Error & PE   & APE \\ \hline
        
        9.00 & 9.00 & 158471 & 237751 & 39.62 & 1.38 & 1.22 & -0.16 & -13.29\% & 13.29\% \\ \hline
9.00 & 9.00 & 174204 & 261182 & 43.54 & 1.51 & 1.38 & -0.13 & -9.18\%  & 9.18\%  \\ \hline
9.00 & 9.00 & 182086 & 272834 & 45.49 & 1.57 & 1.31 & -0.26 & -19.52\% & 19.52\% \\ \hline
9.00 & 9.01 & 192758 & 289429 & 48.22 & 1.66 & 1.70 & 0.03  & 2.06\%   & 2.06\%  \\ \hline
9.00 & 9.01 & 194589 & 291384 & 48.60 & 1.67 & 1.71 & 0.04  & 2.33\%   & 2.33\%  \\ \hline
9.00 & 9.04 & 210728 & 318443 & 52.92 & 1.82 & 1.90 & 0.08  & 4.27\%   & 4.27\%  \\ \hline
9.00 & 9.05 & 218132 & 330068 & 54.82 & 1.88 & 1.96 & 0.08  & 4.13\%   & 4.13\%  \\ \hline
9.00 & 9.05 & 218992 & 330342 & 54.93 & 1.88 & 1.94 & 0.06  & 2.84\%   & 2.84\%  \\ \hline
9.00 & 9.06 & 219841 & 333409 & 55.33 & 1.90 & 1.95 & 0.05  & 2.66\%   & 2.66\%  \\ \hline
9.00 & 9.09 & 236501 & 359469 & 59.60 & 2.04 & 1.80 & -0.24 & -13.16\% & 13.16\% \\ \hline
9.00 & 9.11 & 240700 & 366844 & 60.75 & 2.07 & 1.82 & -0.26 & -14.21\% & 14.21\% \\ \hline
9.00 & 9.11 & 241699 & 368532 & 61.02 & 2.08 & 1.97 & -0.12 & -5.92\%  & 5.92\%  \\ \hline
9.00 & 9.12 & 243667 & 372162 & 61.58 & 2.10 & 2.35 & 0.25  & 10.65\%  & 10.65\% \\ \hline
9.00 & 9.17 & 260945 & 401028 & 66.20 & 2.25 & 2.47 & 0.21  & 8.61\%   & 8.61\%  \\ \hline
9.00 & 9.17 & 262716 & 403213 & 66.59 & 2.27 & 2.43 & 0.16  & 6.65\%   & 6.65\%  \\ \hline
9.00 & 9.36 & 306242 & 483054 & 78.93 & 2.67 & 2.89 & 0.21  & 7.42\%   & 7.42\%  \\ \hline
9.00 & 9.36 & 306564 & 484136 & 79.07 & 2.68 & 2.94 & 0.27  & 9.02\%   & 9.02\%  \\ \hline
9.00 & 9.37 & 308359 & 486893 & 79.53 & 2.69 & 2.93 & 0.24  & 8.03\%   & 8.03\%  \\ \hline
9.00 & 9.44 & 322335 & 515298 & 83.76 & 2.83 & 3.04 & 0.21  & 6.87\%   & 6.87\%  \\ \hline
9.00 & 9.47 & 329852 & 527576 & 85.74 & 2.90 & 3.00 & 0.11  & 3.50\%   & 3.50\%  \\ \hline
    \end{tabular}
    \label{tab:app-gg1-child1}
\end{table}

\begin{table}[!ht]
\centering
  \caption{Validation experiment 2, Scenario 2 results: number of instructions estimated for the abstracted subsystems 2, 3, \& 4 for the simplified $G/G/$ case.}
  \label{tab:app-gg1-child2}
  \footnotesize
    \begin{tabular}{|p{1 cm}|p{1 cm}|p{1 cm}|p{2.5cm}|p{2.5cm}|p{2.5cm}|}
    \hline
    $\theta_{ss2}$ & $\theta_{ss3}$ & $\theta_{ss4}$ & RNI S2 ($I_2=n_2\times(\theta_{ss2}-2))$ & RNI S3 ($I_3 = n_3 \times \theta_{ss3}$) & RNI S4 ($I_4 = n_4\times \theta_{ss4}$) \\
    \hline
      
        9.00 & 9.00 & 9.00 & 509407 & 203749 & 305680 \\ \hline
        9.00 & 9.00 & 9.00 & 559795 & 223977 & 335806 \\ \hline
        9.00 & 9.00 & 9.00 & 584900 & 234110 & 350787 \\ \hline
        9.00 & 9.00 & 9.01 & 619721 & 247832 & 372057 \\ \hline
        9.00 & 9.00 & 9.01 & 624410 & 250186 & 374550 \\ \hline
        9.00 & 9.00 & 9.04 & 678022 & 270936 & 408955 \\ \hline
        9.00 & 9.00 & 9.05 & 701709 & 280456 & 423688 \\ \hline
        9.00 & 9.00 & 9.05 & 703327 & 281561 & 424035 \\ \hline
        9.00 & 9.00 & 9.06 & 708036 & 282653 & 427914 \\ \hline
        9.00 & 9.00 & 9.09 & 760070 & 304073 & 460801 \\ \hline
        9.00 & 9.00 & 9.11 & 774003 & 309471 & 470074 \\ \hline
        9.00 & 9.00 & 9.11 & 777131 & 310756 & 472196 \\ \hline
        9.00 & 9.00 & 9.12 & 783818 & 313287 & 476751 \\ \hline
        9.00 & 9.00 & 9.17 & 838776 & 335500 & 512886 \\ \hline
        9.00 & 9.00 & 9.17 & 843409 & 337778 & 515609 \\ \hline
        9.01 & 9.00 & 9.36 & 986553 & 393740 & 614374 \\ \hline
        9.01 & 9.00 & 9.36 & 987685 & 394154 & 615707 \\ \hline
        9.02 & 9.00 & 9.37 & 992888 & 396462 & 619083 \\ \hline
        9.03 & 9.00 & 9.44 & 1041759 & 414431 & 653798 \\ \hline
        9.04 & 9.00 & 9.47 & 1064131 & 424095 & 668744 \\ \hline
    \end{tabular}
\end{table}

\begin{table}[!ht]
  \centering
  \caption{Validation experiment 2, Scenario 2 results: comparing expected and actual runtime savings for the child $G/G/$ case.}
    \label{tab:app-mgi-child2a}%
  \footnotesize
    \begin{tabular}{|p{1.5 cm}|p{1.5 cm}|p{1.5 cm}|p{1.5 cm}|p{1.5
 cm}|}
    \hline
    Predicted RS & Actual RS & Error & PE   & APE \\
    \hline
       
        3.43 & 3.52 & 0.09  & 2.69\%   & 2.69\%  \\ \hline
3.76 & 3.95 & 0.20  & 4.94\%   & 4.94\%  \\ \hline
3.92 & 3.90 & -0.03 & -0.70\%  & 0.70\%  \\ \hline
4.15 & 4.39 & 0.24  & 5.41\%   & 5.41\%  \\ \hline
4.18 & 4.07 & -0.12 & -2.94\%  & 2.94\%  \\ \hline
4.54 & 4.43 & -0.11 & -2.45\%  & 2.45\%  \\ \hline
4.70 & 4.55 & -0.15 & -3.28\%  & 3.28\%  \\ \hline
4.71 & 4.56 & -0.15 & -3.36\%  & 3.36\%  \\ \hline
4.74 & 4.61 & -0.13 & -2.89\%  & 2.89\%  \\ \hline
5.09 & 4.40 & -0.70 & -15.82\% & 15.82\% \\ \hline
5.19 & 4.49 & -0.70 & -15.54\% & 15.54\% \\ \hline
5.21 & 4.71 & -0.49 & -10.49\% & 10.49\% \\ \hline
5.25 & 4.60 & -0.65 & -14.20\% & 14.20\% \\ \hline
5.63 & 5.56 & -0.06 & -1.13\%  & 1.13\%  \\ \hline
5.66 & 5.67 & 0.01  & 0.18\%   & 0.18\%  \\ \hline
6.64 & 6.69 & 0.05  & 0.80\%   & 0.80\%  \\ \hline
6.65 & 6.66 & 0.01  & 0.20\%   & 0.20\%  \\ \hline
6.68 & 6.76 & 0.08  & 1.16\%   & 1.16\%  \\ \hline
7.02 & 6.93 & -0.08 & -1.22\%  & 1.22\%  \\ \hline
7.17 & 7.06 & -0.11 & -1.55\%  & 1.55\%  \\ \hline
    \end{tabular}
\end{table}

\newpage
\subsection{Details Regarding the Study Code on Github.}
\label{app:github}

The code developed and used for this work can be accessed from its \href{https://github.com/shoaibiocl/ModelSimplificationDES}{GitHub} repository. The repository has the following files:
\begin{enumerate}
    \item ArrivalCount: This corresponds to the first stage of the methodology (discussed in Section \ref{sec:implem}). The code was used to obtain a relationship between the server occupancy values and the average number of instructions per arrival by analysing the simulation trace. 
    The code extracts trace as an output for a 2-stage tandem queueing system (parent model shown in Figure \ref{fig:1}) and for the \textit{1s} and \textit{ms} systems if appropriate modifications are made. 
    \item RunTime: This piece of code was used for the second stage of the methodology to establish a relationship between the average number of instructions and the number of arrivals. The output from the model is recorded in an MS Excel spreadsheet and 
    contains server occupancy values, average number of arrivals, and computational runtime data.
    \item DistributionFitting: The code in this file was used to fit various distributions to LOS data as part of the development of the simplified model, identify the best fitting distribution and estimate its parameters. 
    \item NpKdeFitting: The code in this file was utilised to fit a kernel density estimator (KDE), when no parametric distribution adequately fit the data.
    \item ValidationKlDivergence: The code in this file was used to perform Kullback-Leibler (KL) divergence tests. This compares LOS distributions generated by the simplified model against those generated by the parent model. 
    \item 3StageValidationParentModel: This file contains the code used for validation experiment 2. It is a 3-stage tandem queueing system with two parallel subsystems in the third stage. 
    \item The code for the first validation experiment was a modification of the RunTime file. 
    \item The PHC model code corresponding to validation experiment 3 can be accessed from the link: \href{https://github.com/shoaibiocl/PHC-/blob/main/PHC.py}{PHC code}.  

\end{enumerate}






\end{document}